\documentclass[12pt,preprint,pra,aps,eqsecnum]{revtex4}
\usepackage{amsmath}   

\begin{document} 
\begin{flushright}
 \textbf{CERN-TH/2002-097} \\
\end{flushright}

\title{\vspace*{24pt}
\LARGE\textbf{Theory and application of Fermi
pseudo-potential in one dimension}\\[24pt]} 

\author{\large\textbf{Tai Tsun Wu}\vadjust{\kern10pt}} 

\affiliation{Gordon McKay Laboratory, Harvard University, 
Cambridge, Massachusetts, U.S.A.,}

\affiliation{Theoretical Physics Division, CERN, Geneva,
Switzerland\vadjust{\kern4pt}}

\author{\large\textbf{and\\[16pt] Ming Lun Yu}\vadjust{\kern10pt}}

\affiliation{41019 Pajaro Drive, Fremont, California,
U.S.A.\vadjust{\kern30pt}}
\altaffiliation{Present address: Applied Materials, Inc., 3050 
Bowers Ave., Santa Clara, California, U.S.A.
\vfil
\begin{flushleft}
 \normalsize \textbf{CERN-TH/2002-097} \\
 \textbf{May 2002}
\end{flushleft}
\vskip-\baselineskip\eject}   

\begin{abstract}
\begin{quote}
\hskip1.5em The theory of interaction at one point is developed
for the one-dimensional Schr\"odinger equation.  In analog with the
three-dimensional case, the resulting interaction is referred to as the
Fermi pseudo-potential.  The dominant feature of this one-dimensional
problem comes from the fact that the real line becomes disconnected when
one point is removed.  The general interaction at one point is found to
be the sum of three terms, the well-known delta-function potential and
two Fermi pseudo-potentials, one odd under space reflection and the other
even.  The odd one gives the proper interpretation for the $\delta'(x)$
potential, while the even one is unexpected and more interesting.  Among
the many applications of these Fermi pseudo-potentials, the simplest one
is described.  It consists of a superposition of the delta-function
potential and the even pseudo-potential applied to two-channel
scattering.  This simplest application leads to a model of the quantum
memory, an essential component of any quantum computer.
\end{quote}
\end{abstract} 
\maketitle 
\thispagestyle{empty}
\eject
\section{Introduction}\label{sec:1}

There are several interrelated motivations for the present
investigation.  These are discussed in the following. 

   It was realized several years ago that there are significant
differences between scattering in one channel and scattering in two or more
coupled channels \cite{Khuri}.  For this reason, it may be useful to gain
some experience in dealing with coupled channels in general.

The first question is:  What is the simplest scattering problem in the
case of coupled channels?  Once this simplest problem is understood, it
is reasonable to expect that many of its features hold also for more
general situations.  Clearly, for this simplest problem, the number of
channels should be chosen to be the smallest, namely two, and the number
of spatial dimensions should also be chosen this way, namely one.  Thus
the scattering problem under consideration, in the time-independent case,
deals with the coupled Schr\"odinger equations
\begin{eqnarray}
-{\displaystyle \frac{d^2\psi_1(x)}{dx^2}}+V_{11}(x)\psi_1(x)
+V_{12}(x)\psi_2(x)&=&k^2\psi_1(x),\nonumber\\[8pt]
-{\displaystyle \frac{d^2\psi_2(x)}{dx^2}}+V_{21}(x)\psi_1(x)
+V_{22}(x)\psi_2(x)&=&k^2\psi_2(x),\label{eq:1.1} 
\end{eqnarray}
where the $2\times 2$ matrix potential
\begin{equation}
V(x)=\left[\begin{matrix}
V_{11}(x)\quad &V_{12}(x)\\[3pt]
V_{21}(x)\quad&V_{22}(x)\end{matrix}
\right]\label{eq:1.2}
\end{equation}
is such that it cannot be diagonalized simultaneously for all $x$.

What is the simplest possible choice for this $V(x)$?  In the case of one
channel, the simplest potential is the one that is proportional to the
Dirac delta-function $\delta(x-x_0)$.  This potential is localized at the
one point $x_0$, and the corresponding Schr\"odinger equation is easy to
solve.  For two coupled channels, it is equally desirable to have the
potential localized at one point, say at $x=0$.  However, it is not
allowed to take $V(x)$ to be the product of $\delta(x)$ and a constant
$2\times 2$ matrix because the diagonalization of this constant matrix 
decouples the channels. 

What is needed is therefore another one-dimensional potential that is
localized at one point.  With a linear combination of $\delta(x)$ and
this new potential, the $2\times 2$ matrix $V(x)$ can be easily chosen
such that it cannot be diagonalized and, hence, the two channels do not
decouple.

There are many practical applications of the two-channel scattering
problem in one dimension.  In this paper, let us restrict ourselves to
one such application that is of current interest.  The coupled channels
can be used as a model for quantum memory; a natural approach to
resetting, reading, and writing on a quantum memory is to use scattering
from such a quantum memory.

In trying to find this second potential that is localized at one point,
it is not necessary to study the coupled Schr\"odinger equations
(\ref{eq:1.1}); it is sufficient to return to the simpler case of the
one-channel Schr\"odinger equation
\begin{equation}
\left[-\frac{d^2}{dx^2}+V(x)\right]\psi(x)=k^2\psi(x).\label{eq:1.3}
\end{equation}
A natural first guess for this second potential is the derivative of the
Dirac delta-function, i.e., $\,\delta'(x)$.  However, the presence of
this $\delta'(x)$ term in Eq.\ (\ref{eq:1.3}) implies that the wave
function must be discontinuous at this point $x=0$.  But the product of
$\delta'(x)$ and a function discontinuous at $x=0$ is not well defined. 
Furthermore, even if this $\delta'(x)$ potential is well defined, it is
not suitable for the first application to quantum memory.  The reason for
this will be discussed later in this paper.  

A more powerful method is needed to find this desired potential.  It is
useful here to recall the concept of the Fermi pseudo-potential in three
dimensions, which can be written in the form 
\begin{equation}
\delta^3(r)\,\frac{\partial}{\partial r}\,r,\label{eq:1.4}  
\end{equation}
as given by Blatt and Weisskopf \cite{blattweisskopf}.  The most
far-reaching application of this Fermi pseudo-potential is to the study
of many-body systems, as initiated by Huang and
Yang \cite{huangyang1957}.  For the ground-state energy per particle of a
Bose system of hard spheres, the low-density expansion is known to be
\begin{equation}
4\pi a\rho\left[1+\frac{128}{15\sqrt{\pi}}\,(\rho a^3)^{1/2}
+8\left(\frac{4\pi}{3}-\sqrt{3}\right)\rho a^3\ln(\rho a^3)+O(\rho
a^3)\right].\label{eq:1.5}
\end{equation}
In this expansion, the second term was first obtained by Lee and
Yang \cite{leeyang1957} using the method of binary collision, but the
derivation by Lee, Huang and Yang \cite{leehuangyang1957} using the Fermi
pseudo-potential is somewhat simpler; the third term, which involves the
logarithm, was first obtained by using the Fermi
pseudo-potential \cite{ttwu1959}.  In the derivation of the third term,
it was found inconvenient to use the form (\ref{eq:1.4}), and thus a
limiting process was reintroduced.  This point will be of importance in
this paper.  Thus a great deal is known about the Fermi pseudo-potential
in three dimensions. 

It is a second motivation for this paper to develop the Fermi
pseudo-potential for one-dimen\-sional scattering.  In many cases, once a
theory has been developed for three dimensions, it is straightforward to
repeat the development for one dimension.  In the present case of the
Fermi pseudo-potential, this is not the case.  Furthermore, the result
for one dimension seems qualitatively different from that for three
dimensions. 

For clarity of presentation, this paper is organized into two parts: 
Part~\ref{part:A} for the theory of the Fermi pseudo-potential in one
dimension, and Part~\ref{part:B} for its application to quantum computing. 
Needless to say, these two parts are closely related to each other.  The
sections are numbered consecutively throughout the paper. 
   
\part{\large\textbf{Theory of One-Dimensional Fermi
Pseudo-Potential}}\label{part:A} 

\section{Interaction at One Point}\label{sec:2}

In the absence of $V$, the Hamiltonian of Eq.\ (\ref{eq:1.3}) is
\begin{equation}
H_0=-\frac{d^2}{dx^2}\label{eq:2.1}
\end{equation}
for real $x$, where the right-hand side is suitably interpreted so that
it is self-adjoint.  Let $k$ be purely imaginary; define a real, positive
$\kappa$ by
\begin{equation}
\kappa=-i k.\label{eq:2.2}
\end{equation}
For such a $k$, the Green's function, or resolvent, for this $H_0$
satisfies the differential equation
\begin{equation}
\left(-\frac{d^2}{dx^2}+\kappa^2\right)R_{\kappa}^{(0)}(x,x')=
\delta(x-x'),\label{eq:2.3}
\end{equation}
and is given explicitly by
\begin{equation}
R_{\kappa}^{(0)}(x,x')=(2\kappa)^{-1}e^{-\kappa |x-x'|}.\label{eq:2.4}
\end{equation}

Let a potential $V$ be added to this $H_0$ to give
\begin{equation}
H=H_0+V,\label{eq:2.5}
\end{equation}
which is also self-adjoint.  Again for $\kappa$ positive, the Green's
function, or resolvent, for this $H$ satisfies, similar to Eq.\
(\ref{eq:2.3}),
\begin{equation}
(H+\kappa^2)R_{\kappa}(x,x')=\delta(x-x').\label{eq:2.6}
\end{equation}

The interaction $V$ is said to be at the one point $x_0$ if Eq.\
(\ref{eq:2.6}) implies that Eq.\ (\ref{eq:2.3}), with
$R_{\kappa}^{(0)}(x,x')$ replaced by $R_{\kappa}(x,x')$, is satisfied for
all $x$ except $x=x_0$.

Because of translational symmetry, this $x_0$ is chosen to be 0
throughout this paper.  It is a consequence of Eq.\ (\ref{eq:2.4}) and the
symmetry of the Green's function that this definition of an interaction at
one point implies
\begin{equation}
R_{\kappa}(x,x')=\frac{1}{2\kappa}\left[e^{-\kappa|x-x'|} -f(\kappa;
\mbox{sg}\,x, \mbox{sg}\,x')e^{-\kappa(|x|+|x'|)}\right],\label{eq:2.7}
\end{equation}
where sg$\,x$ and sg$\,x'$ mean the sign of $x$ and the sign of $x'$,
respectively.

Since Eq.\ (\ref{eq:2.7}) is the starting point for the present paper,
this is the appropriate place to add the following comments.

(1) In the present case of one dimension, the real line with the
point $x=0$ removed is not connected.  This is a qualitative difference
between one dimension and higher dimensions. 

(2) It is because of this property that the $f$ in Eq.\ (\ref{eq:2.7})
can depend on the signs of $x$ and $x'$.  

(3) In the three-dimensional case, the Fermi pseudo-potential can be
obtained in the following way:  Take the self-adjoint Hamiltonian
$-\nabla^2$, where $\nabla^2$ is the three-dimensional Laplacian, and
restrict it to functions that are zero at $\textbf{r}=0$; the self-adjoint
extensions \cite{flamand1967} of this restricted operator give the Fermi
pseudo-potential, i.e., these self-adjoint extensions can be written as
the sum of $-\nabla^2$ and (\ref{eq:1.4}) multiplied by a constant.  Such
a procedure applied to the case of one dimension does not give Eq.\
(\ref{eq:2.7}).

It is useful to write out explicitly the $f$ of Eq.\ (\ref{eq:2.7})
as:
\begin{equation}
f(\kappa;\mbox{sg}\,x,\mbox{sg}\,x')=\begin{cases}
f_1(\kappa),\quad&\mbox{for}\ x>0,\ x'>0\\[-2pt] 
f_2(\kappa),&\mbox{for}\ x<0,\ x'>0\\[-2pt]
f_3(\kappa),&\mbox{for}\ x<0,\ x'<0\\[-2pt]
f_4(\kappa),&\mbox{for}\ x>0,\ x'<0\end{cases}
\qquad\label{eq:2.8}
\end{equation}
following the four quadrants in the $x-x'$ plane.  Note that all of these
$f$'s are dimensionless. 

\section{Resolvent Equation}\label{sec:3}

In view of Eq.\ (\ref{eq:2.7}), it is most convenient to study the
resolvent equation in coordinate representation:
\begin{equation}
R_{\kappa_1}(x,x')
-R_{\kappa_2}(x,x')+(\kappa_1^2-\kappa_2^2)\int_{-\infty}^{\infty}
dx''\,R_{\kappa_1}(x,x'')R_{\kappa_2}(x'',x')=0,\label{eq:3.1}
\end{equation}
where $\kappa_1$ and $\kappa_2$ are two values of $\kappa$. 

The substitution of Eq.\ (\ref{eq:2.7}) into this resolvent equation
(\ref{eq:3.1}) gives, after a lengthy calculation,
\begin{eqnarray}
\lefteqn{\frac{1}{\kappa_1-\kappa_2}\,f(\kappa_1;\mbox{sg}\,x,\mbox{sg}
\,x')
+\frac{1}{\kappa_1+\kappa_2}\,f(\kappa_1;\mbox{sg}\,x,-\mbox{sg}\,x')}
\quad\nonumber\\[3pt]
\lefteqn{\mbox{}-\frac{1}{\kappa_1-\kappa_2}\,f(\kappa_2;\mbox{sg}\,x,
\mbox{sg}\,x')+\frac{1}{\kappa_1+\kappa_2}\,f(\kappa_2;-\mbox{sg}\,x,
\mbox{sg}\,x')}\quad\nonumber\\[3pt]
\lefteqn{\mbox{}-\frac{1}{\kappa_1+\kappa_2}\,[f(\kappa_1;\mbox{sg}\,x,-)
f(\kappa_2;-,\mbox{sg}\,x')
+f(\kappa_1;\mbox{sg}\,x,+)f(\kappa_2;+,\mbox{sg}\,x')]}
\quad\nonumber\\[8pt] 
&=&0\hskip4.5in\label{eq:3.2}
\end{eqnarray}
for $\kappa_1\ne \kappa_2$.  This is the resolvent equation for the
interaction at the point $x=0$ as defined in Sec.~\ref{sec:2}. 

Equation (\ref{eq:3.2}) has the following symmetry properties
besides space reflection.

(1) Since $f(\kappa;\mbox{sg}\,x,\mbox{sg}\,x')$ is dimensionless,
there is no scale for $\kappa$.  Thus, Eq.\ (\ref{eq:3.2}) is invariant
under the scale change
\begin{equation}
\kappa_1\to \lambda \kappa_1\quad\mbox{and}\quad \kappa_2\to \lambda
\kappa_2.\label{eq:3.3}
\end{equation}
Note that $\lambda$ is positive since the $\kappa$'s are positive.

(2) There is an additional symmetry
\begin{equation}
f(\kappa;\mbox{sg}\,x,\mbox{sg}\,x')\to -\mbox{sg}\,x\ \mbox{sg}\,x'\
f\left(\frac{1}{\kappa};\mbox{sg}\,x,\mbox{sg}\,x'\right).\label{eq:3.4}
\end{equation}
This discrete symmetry is going to play an important role in this
paper.  In terms of the $f_j(\kappa)$ defined in Eq.\ (\ref{eq:2.8}),
this symmetry is
\begin{displaymath}
\kappa\to\frac{1}{\kappa},
\end{displaymath}
\eject
\vspace*{-2\baselineskip}
\begin{eqnarray}
&f_1\to -f_1,\qquad&f_3\to -f_3,\nonumber\\[4pt]
&f_2\to f_2,\hphantom{-}\qquad &f_4\to f_4.\label{eq:3.5}
\end{eqnarray}

The next task is to solve the resolvent equation (\ref{eq:3.2}) for
$f(\kappa;\mbox{sg}\,x,\mbox{sg}\,x')$.  Since differential equations are
easier to deal with than difference equations, it is convenient to take
the limit $\kappa_1\to \kappa_2$.  In this limit, Eq.\ (\ref{eq:3.2})
reduces to
\begin{eqnarray}
\lefteqn{\frac{d}{d\kappa}\,f(\kappa;\mbox{sg}\,x,\mbox{sg}\,x')
+\frac{1}{2\kappa}\,[f(\kappa;\mbox{sg}\,x,-\mbox{sg}\,x')
+f(\kappa;-\mbox{sg}\,x,\mbox{sg}\,x')]}\quad\nonumber\\[3pt]
\lefteqn{\mbox{}-\frac{1}{2\kappa}\,[f(\kappa;\mbox{sg}\,x,-)
f(\kappa;-,\mbox{sg}\,x') +f(\kappa;\mbox{sg}\,x,+)
f(\kappa;+,\mbox{sg}\,x')]}\quad\nonumber\\[6pt]
&=&0.\hskip4in\label{eq:3.6} 
\end{eqnarray}
In terms of the $f_j(\kappa)$ of Eq.\ (\ref{eq:2.8}), this differential
equation (\ref{eq:3.6}) consists of the following four equations by
taking various signs for $x$ and $x'$:  
\begin{subequations} 
\label{eq:3.7}
\begin{eqnarray}
f'_1(\kappa) +\frac{1}{2\kappa}\,[f_2(\kappa)+f_4(\kappa)]
-\frac{1}{2\kappa}\,[f_2(\kappa)f_4(\kappa)+f_1(\kappa)^2]&=&0,
\label{eq:3.7a}\\[4pt]
f'_2(\kappa)+\frac{1}{2\kappa}\,[f_1(\kappa)+f_3(\kappa)]-
\frac{1}{2\kappa}\,
[f_2(\kappa)f_3(\kappa)+f_1(\kappa)f_2(\kappa)]&=&0,\label{eq:3.7b}\\[4pt]
f'_3(\kappa)+\frac{1}{2\kappa}\,[f_2(\kappa)+f_4(\kappa)]
-\frac{1}{2\kappa}\,
[f_3(\kappa)^2+f_2(\kappa)f_4(\kappa)]&=&0,\label{eq:3.7c}\\[4pt] 
f'_4(\kappa)+\frac{1}{2\kappa}\,[f_1(\kappa)+f_3(\kappa)]
-\frac{1}{2\kappa}\,
[f_3(\kappa)f_4(\kappa)+f_1(\kappa)f_4(\kappa)]&=&0.\label{eq:3.7d}
\end{eqnarray}
\end{subequations} 

An examination of these four differential equations shows the important
role played by the combination $f_1(\kappa)+f_3(\kappa)$, which appears
twice in Eq.\ (\ref{eq:3.7b}) and twice in Eq.\ (\ref{eq:3.7d}). 
Define $F(\kappa)$ up to an additive constant by
\begin{equation}
F'(\kappa)=\frac{1}{2\kappa}\,[f_1(\kappa)+f_3(\kappa)].\label{eq:3.8}
\end{equation}
In terms of this $F(\kappa)$, Eqs.\ (\ref{eq:3.7b}) and (\ref{eq:3.7d})
take the form
\begin{equation}
f'_2(\kappa)+F'(\kappa)-f_2(\kappa)F'(\kappa)=0;\qquad
f'_4(\kappa)+F'(\kappa)-f_4(\kappa)F'(\kappa)=0.\label{eq:3.9} 
\end{equation}  
Integration of Eqs.\ (\ref{eq:3.9}) gives $f_2(\kappa)$ and $f_4(\kappa)$
in terms of $F(\kappa)$: 
\begin{equation}
f_2(\kappa)=1+c_2e^{F(\kappa)};\qquad
f_4(\kappa)=1+c_4e^{F(\kappa)},\label{eq:3.10} 
\end{equation} 
where $c_2$ and $c_4$ are two arbitrary constants of integration. 
Similarly, subtracting Eq.\ (\ref{eq:3.7c}) from Eq.\ (\ref{eq:3.7a})
gives
\begin{equation}
\frac{d}{d\kappa}\,[f_1(\kappa)-f_3(\kappa)]=
\frac{1}{2\kappa}\,[f_1(\kappa)^2-f_3(\kappa)^2]=F'(\kappa)[f_1(\kappa)
-f_3(\kappa)],\label{eq:3.11}
\end{equation}  
or
\begin{equation}
f_1(\kappa)-f_3(\kappa)=2c_3e^{F(\kappa)},\label{eq:3.12}
\end{equation}
where $c_3$ is another arbitrary constant of integration.

It remains to determine $F(\kappa)$, which satisfies the second-order
ordinary differential equation obtained from adding Eqs.\ (\ref{eq:3.7a})
and (\ref{eq:3.7c}):
\begin{equation}
2\kappa\,\frac{d}{d\kappa}\,\kappa\,\frac{d}{d\kappa}\,F(\kappa)=
\left[\kappa\,\frac{d}{d\kappa}\,F(\kappa)\right]^2
-1+(c_3^2+c_2c_4)e^{2F(\kappa)}.\label{eq:3.13}
\end{equation}
The solution of this equation is straightforward but somewhat lengthy and
is thus relegated to Appendix~A.

The results are as follows:
\begin{subequations}
\label{eq:3.14}
\begin{eqnarray}
f_1(\kappa)&=&\frac{-\sqrt{c_3^2+c_2c_4-c_1^2}\,[c_0\kappa
+(c_0\kappa)^{-1}]-2c_3}{\sqrt{c_3^2+c_2c_4-c_1^2}\,[c_0\kappa
-(c_0\kappa)^{-1}]+2c_1},\label{eq:3.14a}\\[4pt]
f_2(\kappa)&=&1-\frac{2c_2}{\sqrt{c_3^2+c_2c_4-c_1^2}\,[c_0\kappa
-(c_0\kappa)^{-1}]+2c_1},\label{eq:3.14b}\\[4pt]
f_3(\kappa)&=&\frac{-\sqrt{c_3^2+c_2c_4-c_1^2}\,[c_0\kappa
+(c_0\kappa)^{-1}]+2c_3}{\sqrt{c_3^2 +c_2c_4 -c_1^2}\, [c_0\kappa
-(c_0\kappa)^{-1}]+2c_1},\label{eq:3.14c}\\[4pt]
f_4(\kappa)&=&1-\frac{2c_4}{\sqrt{c_3^2 +c_2c_4 -c_1^2}\, [c_0\kappa
-(c_0\kappa)^{-1}]+2c_1},\label{eq:3.14d}
\end{eqnarray}
\end{subequations}
when
\begin{equation}
c_3^2 +c_2c_4 -c_1^2>0.\label{eq:3.15} 
\end{equation}
Similarly,
\begin{subequations}
\label{eq:3.16} 
\begin{eqnarray}
f_1(\kappa)&=&\frac{-\sqrt{c_1^2 -c_3^2 -c_2c_4}\,[c_0\kappa
-(c_0\kappa)^{-1}] -2c_3}{\sqrt{c_1^2 -c_3^2 -c_2c_4}\,[c_0\kappa
-(c_0\kappa)^{-1}]+2c_1},\label{eq:3.16a}\\[4pt]
f_2(\kappa)&=&1-\frac{2c_2}{\sqrt{c_1^2-c_3^2 -c_2c_4}\,[c_0\kappa
-(c_0\kappa)^{-1}]+2c_1},\label{eq:3.16b}\\[4pt]
f_3(\kappa)&=&\frac{-\sqrt{c_1^2 -c_3^2 -c_2c_4}\,[c_0\kappa
-(c_0\kappa)^{-1}]+2c_3}{\sqrt{c_1^2 -c_3^2 -c_2c_4}\,[c_0\kappa
-(c_0\kappa)^{-1}]+2c_1},\label{eq:3.16c}\\[4pt]
f_4(\kappa)&=&1-\frac{2c_4}{\sqrt{c_1^2 -c_3^2 -c_2c_4}\,[c_0\kappa
-(c_0\kappa)^{-1}]+2c_1},\label{eq:3.16d}
\end{eqnarray}
\end{subequations} 
when
\begin{equation}
c_3^2 +c_2c_4-c_1^2 <0.\label{eq:3.17}
\end{equation}
In Eqs.\ (\ref{eq:3.14}) and (\ref{eq:3.16}), $\,c_0>0$ and all square
roots are also positive.  Although there are five constants $c_0$, $c_1$,
$c_2$, $c_3$ and $c_4$, effectively there are four because all the
quantities do not change under
\begin{equation}
c_j\to \lambda c_j\label{eq:3.18}
\end{equation}
for $j=1$, 2, 3, 4, and $\lambda>0$. 

It remains to discuss briefly the case
\begin{equation}
c_3^2+c_2c_4 -c_1^2=0.\label{eq:3.19}
\end{equation}
This case, which was in fact the first case worked out and also the most
important one as discussed in Sec.\ \ref{sec:4}, can be recovered by
taking the limit $c_3^2+c_2c_4-c_1^2\to 0$ together with either $c_0\to
0$ or $c_0\to\infty$.  These two limiting cases are to be considered
separately.  For definiteness, they are applied to Eqs.\ (\ref{eq:3.16}).

(1) $c_3^2+c_2c_4 -c_1^2\to 0$ and $c_0\to 0$, such that
\begin{equation}
\gamma = c_0^{-1}\sqrt{c_1^2-c_3^2-c_2c_4}\label{eq:3.20}
\end{equation}
is fixed; $\,\gamma>0$.  In this limit,
\begin{subequations} 
\label{eq:3.21}
\begin{eqnarray}
f_1(\kappa)&=& \frac{\gamma -2c_3\kappa}{\gamma
+2c_1\kappa},\label{eq:3.21a}\\[4pt]
f_2(\kappa)&=&1-\frac{2c_2\kappa}{\gamma
+2c_1\kappa},\label{eq:3.21b}\\[4pt]
f_3(\kappa)&=&\frac{\gamma +2c_3\kappa}{\gamma
+2c_1\kappa},\label{eq:3.21c}\\[4pt]
f_4(\kappa)&=&1-\frac{2c_4\kappa}{\gamma +2c_1\kappa}.\label{eq:3.21d}
\end{eqnarray}
\end{subequations} 

(2) $c_3^2+c_2c_4 -c_1^2\to 0$ and $c_0\to\infty$, such that
\begin{equation}
\gamma = c_0\sqrt{c_1^2-c_3^2-c_2c_4}\label{eq:3.22}
\end{equation}
is fixed; $\,\gamma>0$.  In this limit,
\begin{subequations}
\label{eq:3.23}
\begin{eqnarray}
f_1(\kappa)&=&\frac{-\gamma\kappa-2c_3}{\gamma\kappa+2c_1},
\label{eq:3.23a}\\[4pt] 
f_2(\kappa)&=&1-\frac{2c_2}{\gamma\kappa
+2c_1},\label{eq:3.23b}\\[4pt] 
f_3(\kappa)&=&\frac{-\gamma\kappa
+2c_3}{\gamma\kappa +2c_1},\label{eq:3.23c}\\[4pt]
f_4(\kappa)&=&1-\frac{2c_4}{\gamma\kappa +2c_1}.\label{eq:3.23d}
\end{eqnarray}
\end{subequations}
Note that Eqs.\ (\ref{eq:3.21}) and (\ref{eq:3.23}) are related by the
discrete symmetry (\ref{eq:3.5}) provided that the sign of $c_3$ is
reversed.  The same results also follow from Eqs.\ (\ref{eq:3.14}). 

\section{Interaction Potentials}\label{sec:4}

Naively, one would expect it to be straightforward to determine the
potential when the Green's function (resolvent) is known.  It does not
turn out to be so straightforward, and this section is devoted to
solving this problem.

The substitution of Eqs.\ (\ref{eq:2.1}) and (\ref{eq:2.5}) into Eq.\
(\ref{eq:2.6}) gives
\begin{equation}
V(x)R_{\kappa}(x,x')=\delta(x-x')-\left(-\frac{d^2}{dx^2}+\kappa^2\right)
R_{\kappa}(x,x'),\label{eq:4.1}
\end{equation}
or
\begin{equation}
V(x)R_{\kappa}(x,x')=-\left(-\frac{d^2}{dx^2}+\kappa^2\right)
[R_{\kappa}(x,x') -R_{\kappa}^{(0)}(x,x')],\label{eq:4.2}
\end{equation}
with the last term defined by Eq.\ (\ref{eq:2.3}) or Eq.\
(\ref{eq:2.4}).  This should determine $V(x)$; that this $V(x)$ does not
depend on $\kappa$ is a consequence of $R_{\kappa}(x,x')$ satisfying the
resolvent equation.

More generally, the left-hand side of Eq.\ (\ref{eq:4.2}) may be an
integral, and this equation takes the form
\begin{equation}
\int_{-\infty}^{\infty}
dx''\,V(x,x'')R_{\kappa}(x'',x')=-\left(-\frac{d^2}{dx^2}+\kappa^2\right)
[R_{\kappa}(x,x') -R_{\kappa}^{(0)}(x,x')].\label{eq:4.3}
\end{equation}
If $V(x)$ exists, then
\begin{equation}
V(x,x')=V(x)\delta(x-x').\label{eq:4.4}
\end{equation} 

Since Eq.\ (\ref{eq:4.2}) is simpler than Eq.\ (\ref{eq:4.3}), it is
useful to study Eq.\ (\ref{eq:4.2}) first even though it is less
general.  The substitution of Eq.\ (\ref{eq:2.7}) into Eq.\ (\ref{eq:4.2})
gives
\begin{equation}
V(x)R_{\kappa}(x,x')=\frac{1}{2\kappa}\left(-\frac{d^2}{dx^2}+\kappa^2
\right)
f(\kappa;\mbox{sg}\,x,\mbox{sg}\,x')e^{-\kappa(|x|+|x'|)}.\label{eq:4.5}
\end{equation}
The difficulty is to give a proper interpretation to this equation.  The
right-hand side contains a term
\begin{equation}
f(\kappa;\mbox{sg}\,x,\mbox{sg}\,x')\delta(x)e^{-\kappa|x'|},\label{eq:4.6}
\end{equation}
obtained by applying the differential operator to the exponential.  As
seen from Eqs.\ (\ref{eq:3.14}) for example, this expression
(\ref{eq:4.6}) is in general the product of $\delta(x)$ and a function
discontinuous at $x=0$.  The only reasonable interpretation of such a
product is 
\begin{equation}\alpha(x)\delta(x)=\frac{1}{2}\left[\left(\lim_{x\to 0+}
+\lim_{x\to 0-}\right)\alpha(x)\right]\delta(x).\label{eq:4.7}
\end{equation}

As mentioned above, there are effectively four parameters in the
solutions as given by either Eqs.\ (\ref{eq:3.14}) or Eqs.\
(\ref{eq:3.16}).  Since the symmetry of the Green's function implies that
$c_2=c_4$, the number of parameters is reduced by 1.  Therefore, the
potential $V(x,x')$ depends on {\it three} parameters.  That there are
three parameters instead of two is a major surprise, and this fact is to
play a central role in Part~\ref{part:B} of this paper, where this
interaction potential is applied to study certain aspects of quantum
computing. 

The three pieces of $V(x,x')$ are of different levels of complication. 
They are studied in the following three subsections.

\subsection{$\pmb{\delta(x)}$ potential}\label{subsec:a}

The simplest piece is the well-known $\delta(x)$ potential, where
\begin{equation}
V_1(x)=g_1\delta(x),\label{eq:4.8}  
\end{equation}
and
\begin{equation}
V_1(x,x')=g_1\delta(x)\delta(x-x'),\label{eq:4.9}
\end{equation}
or equivalently
\begin{equation}
V_1(x,x')=g_1\delta(x)\delta(x').\label{eq:4.10}
\end{equation}
For this potential, the differential equation (\ref{eq:2.6}) is well
defined.  Its solution is
\begin{equation}
R_{\kappa}(x,x')=\frac{1}{2\kappa}\left[e^{-\kappa|x-x'|}
-\frac{g_1}{2\kappa +g_1}\,e^{-\kappa(|x|+|x'|)}\right].\label{eq:4.11}
\end{equation}
Comparison with Eq.\ (\ref{eq:2.7}) gives
\begin{equation}
f(\kappa;\mbox{sg}\,x,\mbox{sg}\,x')=\frac{g_1}{2\kappa
+g_1},\label{eq:4.12}
\end{equation}
independent of the signs of $x$ and $x'$.  By Eq.\ (\ref{eq:2.8}), this is
\begin{equation}
f_1(\kappa)=f_2(\kappa)=f_3(\kappa)=f_4(\kappa)=
\frac{g_1}{2\kappa+g_1}.\label{eq:4.13}
\end{equation} 
This is a special case of Eqs.\ (\ref{eq:3.21}) with
\begin{equation}
c_1=c_2=c_4=\frac{\gamma}{g_1}\quad\mbox{and}\quad c_3=0.\label{eq:4.14}
\end{equation}

It is instructive to recover Eq.\ (\ref{eq:4.8}) from Eq.\
(\ref{eq:4.12}) by using Eq.\ (\ref{eq:4.5}).  Since
$f(\kappa;\mbox{sg}\,x,\mbox{sg}\,x')$ does not depend on the signs of
$x$ and $x'$, Eq.\ (\ref{eq:4.5}) gives
\begin{eqnarray}
V(x)&=&\left[\frac{g_1}{2\kappa+g_1}\,\delta(x)e^{-\kappa|x'|}\right]
\biggl/\left\{
\frac{1}{2\kappa}\left[e^{-\kappa|x-x'|}
-\frac{g_1}{2\kappa+g_1}\,e^{-\kappa(|x|+|x'|)}\right]
\right\}\nonumber\\[6pt] &=&g_1\delta(x).\label{eq:4.15}
\end{eqnarray}
This $V(x)$ is even in~$x$.

\subsection{$\pmb{\delta'(x)}$ potential}\label{subsec:b} 

As already discussed in Sec.\ \ref{sec:1}, the potential $\delta'(x)$ is
not acceptable because, in the Schr\"odinger equation, there is a product
of $\delta'(x)$ and a function discontinuous at $x=0$.  While
$\delta(x)$ is a potential, $\,\delta'(x)$ has to be understood as a
Fermi pseudo-potential in much the same way as the expression
(\ref{eq:1.4}) in three dimensions.

Since $\delta'(x)$ is odd in $x$, the resolvent must satisfy
\begin{equation}
f_1(\kappa)=-f_3(\kappa)\quad\mbox{and}\quad
f_2(\kappa)=f_4(\kappa).\label{eq:4.16}
\end{equation}
Next, consider the {\it formal} Eq.\ (\ref{eq:2.6}) with this
$\delta'(x)$ potential
\begin{equation}
\left[-\frac{d^2}{dx^2}+g_2\delta'(x)+\kappa^2\right]R_{\kappa}(x,x')
=\delta(x-x').\label{eq:4.17}
\end{equation}
Since every term on the left-hand side is of dimension $x^{-2}$ times
that of $R_{\kappa}(x,x')$, the resolvent $R_{\kappa}(x,x')$ must be of
the form
\begin{displaymath}
\kappa^{-1}\ \mbox{function\ of\ }\kappa x\ \mbox{and}\ \kappa
x'.
\end{displaymath}
A comparison with Eq.\ (\ref{eq:2.7}) then shows that
$f(\kappa;\mbox{sg}\,x,\mbox{sg}\,x')$ is independent of $\kappa$.  This
is satisfied with
\begin{equation}
f_1(\kappa)=\alpha,\quad f_3(\kappa)=-\alpha,\quad\mbox{and}\quad
f_2(\kappa)=f_4(\kappa)=-\beta,\label{eq:4.18}
\end{equation}
with
\begin{equation}
2\beta+\alpha^2+\beta^2=0.\label{eq:4.19}
\end{equation}

With the resolvent known, it is now possible to define the Fermi
pseudo-potential
\begin{equation}
V_2(x)=g_2\delta'_p(x).\label{eq:4.20}
\end{equation}
Omitting the argument $\kappa$ in $f$, Eq.\ (\ref{eq:4.5}) takes the form
\begin{equation}
g_2\delta'_p(x)R_{\kappa}(x,x')=\frac{1}{2\kappa}\left(-\frac{d^2}{dx^2}
+\kappa^2\right)f(\mbox{sg}\,x,\mbox{sg}\,x')
e^{-\kappa(|x|+|x'|)}.\label{eq:4.21}
\end{equation}
Let $x'$ be positive.  Then
\begin{equation}
\left(-\frac{d^2}{dx^2}+\kappa^2\right)f(\mbox{sg}\,x,\mbox{sg}\,x')
e^{-\kappa|x|}
=-(f_1-f_2)\delta'(x)+\kappa(f_1+f_2)\delta(x).\label{eq:4.22}
\end{equation}
This expression has not only a $\delta'(x)$ term, but also a $\delta(x)$
term.  For the left-hand side of Eq.\ (\ref{eq:4.21}), it is necessary
to evaluate, using Eq.\ (\ref{eq:4.7}),
\begin{eqnarray}
R_{\kappa}(x,x')\Bigr|_{x=0}&=&\left[e^{-\kappa(x'-x)}
-f(\mbox{sg}\,x,+)e^{-\kappa(|x|+x')}\right]_{x=0}\nonumber\\[4pt]
&=&e^{-\kappa x'}\left[1-\frac{1}{2}\left(\lim_{x\to 0+}+\lim_{x\to
0-}\right)f(\mbox{sg}\,x,+)e^{-\kappa|x|}\right]\nonumber\\[4pt]
&=&e^{-\kappa x'}\left[1-\frac{1}{2}\,(f_1+f_2)\right]\label{eq:4.23}
\end{eqnarray}
and
\begin{eqnarray}
\lefteqn{\frac{d}{dx}\,\left[e^{-\kappa(x'-x)}-f(\mbox{sg}\,x,+)e^{-\kappa
(|x|+x')}\right]_{x=0}}\quad\nonumber\\[4pt]
&=&\kappa\left[e^{-\kappa(x'-x)}+(\mbox{sg}\,x)f(\mbox{sg}\,x,+)e^{-\kappa
(|x|+x')}\right]_{x=0}\nonumber\\[4pt]   
&=&\kappa e^{-\kappa x'}\left[1+\frac{1}{2}\left(\lim_{x\to
0+}+ \lim_{x\to 0-}\right)(\mbox{sg}\,x)f(\mbox{sg}\,x,+)
e^{-\kappa|x|}\right]\nonumber\\[4pt]
&=&\kappa e^{-\kappa x'}\left[1+\frac{1}{2}\,(f_1-f_2)\right].
\label{eq:4.24}
\end{eqnarray}

Suppose the $\delta'_p(x)$ on the left-hand side of Eq.\ (\ref{eq:4.21})
is replaced by $\delta'(x)$.  Then a comparison of Eq.\ (\ref{eq:4.22})
with Eqs.\ (\ref{eq:4.23}) and (\ref{eq:4.24}) gives
\begin{equation}
g_2\left[1-\frac{1}{2}\,(f_1+f_2)\right]=f_1-f_2;\qquad
g_2\left[1+\frac{1}{2}\,(f_1-f_2)\right]=f_1+f_2,\label{eq:4.25}
\end{equation} 
where the identity $x\delta'(x)=-\delta(x)$ has been used.  These are the
conditions for $x'>0$; similar conditions for $x'<0$ are
\begin{equation}
g_2\left[1-\frac{1}{2}\,(f_4+f_3)\right]=f_4-f_3;\qquad
g_2\left[-1+\frac{1}{2}\,(f_4-f_3)\right]=f_4+f_3.\label{eq:4.26}
\end{equation}
Solving Eqs.\ (\ref{eq:4.25}) and (\ref{eq:4.26}) gives
\begin{equation} 
f_1=\frac{g_2}{1+g_2^2/4};\qquad f_3=\frac{-g_2}{1+g_2^2/4};\qquad
f_2=f_4=\frac{g_2^2/2}{1+g_2^2/4}.\label{eq:4.27}
\end{equation}
This is consistent with the previous Eqs.\ (\ref{eq:4.18}) and
(\ref{eq:4.19}).

Where is the difficulty explained in Sec.\ \ref{sec:1}?  Another way of
asking the same question is:  How does $\delta'_p(x)$ differ from
$\delta'(x)$?

The answer is to be found in the first step of Eq.\ (\ref{eq:4.24}).  In
differentiating the quantity on the left-hand side of Eq.\
(\ref{eq:4.24}), the factor $f(\mbox{sg}\,x,+)$ is not differentiated. 
In other words, the term with $(d/dx) f(\mbox{sg}\,x,+)$ has been
omitted; if this term were not omitted, there would be a $\delta(x)$,
precisely the difficulty explained in Sec.~\ref{sec:1}.

The situation is therefore entirely similar to the Fermi pseudo-potential
in three dimensions, where the operator (\ref{eq:1.4}) performs the
function of removing a term proportional to $1/r$.  Here, what
$\delta'_p(x)$ does is 
\begin{equation}
\delta'_p(x)g(x)=\delta'(x)\tilde g(x),\label{eq:4.28}
\end{equation}
where
\begin{equation}
\tilde g(x) = \begin{cases}
{\displaystyle g(x)-\lim_{x\to0+} g(x),}\quad&\mbox{for\ }x>0,\\[3pt]
{\displaystyle g(x)-\lim_{x\to 0-}g(x),}&\mbox{for\ }x<0.
\end{cases}\label{eq:4.29}
\end{equation}
This removes the discontinuity of $g(x)$ at $x=0$, which is precisely what
is needed. 

\subsection{Third potential}\label{subsec:c}

At the beginning of this investigation it was thought that, in one
dimension, there was one potential (subsection \ref{subsec:a}) and one
pseudo-potential (subsection \ref{subsec:b}).  But the detailed analysis
of the resolvent equation in Sec.\ \ref{sec:3} shows that there are three
independent parameters in the resolvent, and hence there is an
independent third potential, or a second pseudo-potential in one
dimension.

This third potential is most easily understood through the discrete
symmetry (\ref{eq:3.4}).  Let $c=2/g_1$.  Then, from Eq.\ (\ref{eq:4.13}),
the resolvent for the $\delta(x)$ potential is given by
\begin{equation}
f_1(\kappa)=f_2(\kappa)=f_3(\kappa)=f_4(\kappa)=\frac{1}{1+c\kappa}.
\label{eq:4.30}
\end{equation}
Application of the discrete symmetry (\ref{eq:3.4}) to Eq.\
(\ref{eq:4.30}) gives the result that the resolvent for the third potential
is expressed by
\begin{equation}
f_1(\kappa)=f_3(\kappa)=\frac{-\kappa}{\kappa+c};\qquad
f_2(\kappa)=f_4(\kappa)=\frac{\kappa}{\kappa+c}.\label{eq:4.31}
\end{equation}
In particular, similar to the resolvent of the first pseudo-potential as
given by Eq.\ (\ref{eq:4.27}), this resolvent is also not continuous. 
The relation between the resolvents as expressed by Eqs.\ (\ref{eq:4.30})
and (\ref{eq:4.31}) is a special case of that between Eqs.\
(\ref{eq:3.21}) and (\ref{eq:3.23}). 

It remains to determine the potential, or more precisely the
pseudo-potential, from Eq.\ (\ref{eq:4.31}), which can be written more
succinctly as
\begin{equation}
f(\kappa;\mbox{sg}\,x,\mbox{sg}\,x')=-\frac{\kappa}{\kappa+c}\,
\mbox{sg}\,x\
\mbox{sg}\,x'.\label{eq:4.32}
\end{equation}
Therefore, for the present case of the third potential $V_3(x,x')$, Eq.\
(\ref{eq:4.3}) takes the form
\begin{eqnarray}
\lefteqn{\int_{-\infty}^{\infty}dx''\,V_3(x,x'')\left[e^{-\kappa|x''-x'|}
+\frac{\kappa}{\kappa +c}\,\mbox{sg}\,x''\ \mbox{sg}\,x'\
e^{-\kappa(|x''|+|x'|)}\right]}\quad\nonumber\\[4pt]
&=&\frac{\kappa}{\kappa
+c}\left(-\frac{d^2}{dx^2}+\kappa^2\right)\left[\mbox{sg}\,x\ 
\mbox{sg}\,x'\
e^{-\kappa(|x|+|x'|)}\right].\hskip.7in\label{eq:4.33} 
\end{eqnarray}
Differentiation of the right-hand side gives
\begin{displaymath}
\frac{d}{dx}\,\left[\mbox{sg}\,x\
e^{-\kappa|x|}\right]=2\delta(x)-\kappa e^{-\kappa|x|},
\end{displaymath}
and hence
\begin{equation}
\left(-\frac{d^2}{dx^2}+\kappa^2\right)\left[\mbox{sg}\,x\
e^{-\kappa|x|}\right] =-2\delta'(x).\label{eq:4.34}
\end{equation}
Thus Eq.\ (\ref{eq:4.33}) for $V_3(x,x')$ is explicitly 
\begin{eqnarray}
\lefteqn{\int_{-\infty}^{\infty}dx''\,V_3(x,x'')\left[e^{-\kappa|x''-x'|}
+\frac{\kappa}{\kappa+c}\,\mbox{sg}\,x''\ \mbox{sg}\,x'\ e^{-\kappa
(|x''|+|x'|)}\right]}\quad\nonumber\\[4pt]
&=&-\frac{2\kappa}{\kappa+c}\,\delta'(x)\,\mbox{sg}\,x'\
e^{-\kappa|x'|}.\hskip2.25in\label{eq:4.35}
\end{eqnarray}
The task is to make sense of this equation and to determine $V_3(x,x')$. 

That the resolvent equation is satisfied means that Eq.\ (\ref{eq:4.35}),
properly understood, does lead to a $V_3(x,x')$.  By Eqs.\
(\ref{eq:4.28}) and (\ref{eq:4.29}), the $\delta'(x)$ on the right-hand
side of Eq.\ (\ref{eq:4.35}) can be replaced by $\delta'_p(x)$, because it
is not multiplied by a discontinuous function of~$x$.  Therefore,
$\,V_3(x,x')$ is expected to be proportional to $\delta'_p(x)$; 
that $\delta'_p(x)$ is used instead of $\delta'(x)$ is due to the
development in subsection \ref{subsec:b}.  With these considerations, an
examination of Eq.\ (\ref{eq:4.35}) indicates that
\begin{equation}
V_3(x,x')=g_3\delta'_p(x)\delta'_p(x').\label{eq:4.36}
\end{equation}
See also Eq.\ (\ref{eq:4.20}).

It remains to substitute Eq.\ (\ref{eq:4.36}) into Eq.\ (\ref{eq:4.35})
to find the relation between the two constants $g_3$ and $c$:
\begin{eqnarray}
\lefteqn{g_3\int_{-\infty}^{\infty}dx''\,\delta'_p(x'')\left[e^{-\kappa
|x''-x'|} +\frac{\kappa}{\kappa+c}\,\mbox{sg}\,x''\ \mbox{sg}\,x'\
e^{-\kappa(|x''|+|x'|)}\right]}\quad\nonumber\\[4pt]
&=&-\frac{2\kappa}{\kappa+c}\,\mbox{sg}\,x'\
e^{-\kappa|x'|}.\hskip2.5in\label{eq:4.37}
\end{eqnarray}
The evaluation of the first integral is straightforward because
$e^{-\kappa|x''-x'|}$ is continuous: 
\begin{eqnarray}
\int_{-\infty}^{\infty}dx''\,\delta'_p(x'')e^{-\kappa|x''-x'|}
&=&\int_{-\infty}^{\infty}dx''\,\delta'(x'')e^{-\kappa|x''-x'|}
\nonumber\\[4pt]
&=&-\kappa\,\mbox{sg}\,x'\ e^{-\kappa|x'|}.\hskip2.25in\label{eq:4.38}
\end{eqnarray} 
After the removal of the common factor
\begin{displaymath}
\kappa\,\mbox{sg}\,x'\ e^{-\kappa|x'|},
\end{displaymath}
Eq.\ (\ref{eq:4.37}) reduces to
\begin{equation}
g_3\left[-1+\frac{1}{\kappa+c}\int_{-\infty}^{\infty}dx''\,\delta'_p(x'')\,
\mbox{sg}\,x''\ e^{-\kappa|x''|}\right]=-\frac{2}{\kappa+c}.\label{eq:4.39}
\end{equation}  
This integral can be evaluated using Eqs.\ (\ref{eq:4.28}) and
(\ref{eq:4.29}):
\begin{equation}
\int_{-\infty}^{\infty}dx''\,\delta'_p(x'')\,\mbox{sg}\,x''\ e^{-\kappa
|x''|}=-\int_{-\infty}^{\infty}dx''\,\delta(x'')\,\mbox{sg}\,x''\
\frac{d}{dx''}\,e^{-\kappa|x''|}=\kappa.\label{eq:4.40}
\end{equation}
Therefore, Eq.\ (\ref{eq:4.39}) is simply
\begin{equation}
g_3\left[-1+\frac{\kappa}{\kappa+c}\right]=-\frac{2}{\kappa+c}\qquad
\mbox{or}
\qquad g_3=\frac{2}{c}.\label{eq:4.41}
\end{equation}
This is the desired relation.

It is merely a matter of terminology whether this pseudo-potential
$V_3(x,x')$ as given by Eq.\ (\ref{eq:4.36}) is called a local potential
or not.

In summary, the three potentials $V_1$, $\,V_2$, and $V_3$ are given by
Eqs.\ (\ref{eq:4.8}), (\ref{eq:4.20}), and (\ref{eq:4.36}).  Thus the
most general Fermi pseudo-potential for the interaction at one point in
one dimension is
\begin{eqnarray}
V(x,x')&=&V_1(x,x') +V_2(x,x')+V_3(x,x')\nonumber\\[4pt]
&=&g_1\delta(x)\delta(x-x')+g_2\delta'_p(x)\delta(x-x')+g_3\delta'_p(x)
\delta'_p(x').\label{eq:4.42}
\end{eqnarray}

From the above experience of working with $\delta'_p(x)$ and the fact
that the product of $\delta'(x)$ and a function discontinuous at $x=0$
is not meaningful, from here on the convention will be adopted that
$\delta'(x)$ always means $\delta'_p(x)$.  With this convention, Eq.\
(\ref{eq:4.42}) is written as
\begin{equation}
V(x,x')=g_1\delta(x)\delta(x-x')
+g_2\delta'(x)\delta(x-x')+g_3\delta'(x)\delta'(x').\label{eq:4.43}
\end{equation}
Equation (\ref{eq:4.43}) can be rewritten in a prettier form as
follows.  Since
\begin{displaymath}
\delta(x)\delta(x-x')=\delta(x)\delta(x')
\end{displaymath}
and
\begin{eqnarray}
\delta'(x)\delta(x-x')&=&\delta'(x)\delta(x'-x)\nonumber\\[4pt]
&=&\delta'(x)[\delta(x')-x\delta'
(x')]\nonumber\\[4pt]
&=&\delta'(x)\delta(x')+\delta(x)\delta'(x'),\label{eq:4.44}
\end{eqnarray} 
where use has been made of the identity $\delta'(x)x=-\delta(x)$, a
general Fermi pseudo-potential (\ref{eq:4.43}) can be written as 
\begin{equation}
V(x,x')=g_1\delta(x)\delta(x')+g_2[\delta'(x)\delta(x')+\delta(x)\delta'(x')]
+g_3\delta'(x)\delta'(x').\label{eq:4.45}
\end{equation}
As already mentioned, the first and last terms are even while the middle
term is odd.  That is, under space inversion 
\begin{equation}
x\to -x\quad\mbox{and}\quad x'\to -x',\label{eq:4.46}
\end{equation}
the coupling constants transform as 
\begin{equation}
g_1\to g_1;\qquad g_2\to -g_2;\qquad g_3\to g_3.\label{eq:4.47}
\end{equation}

\section{Solving the Schr\"odinger Equation}\label{sec:5}

In applying the Fermi pseudo-potential to various problems, such as the
one to be treated in Part~\ref{part:B} of this paper, the resolvent
equation is difficult to use and it is much more convenient to employ the
prescription of Sec.\ \ref{sec:4} to the Schr\"odinger equation.

This section is devoted to studying the equation
\begin{equation}
\left(-\frac{d^2}{dx^2}+\kappa^2\right)R_{\kappa}(x,x')
+\int_{-\infty}^{\infty}dx''\,V(x',x'')R_{\kappa}(x'',x')
=\delta(x-x'),\label{eq:5.1}
\end{equation}
where $V(x,x')$ is the Fermi pseudo-potential as given by Eq.\
(\ref{eq:4.42}).  On the one hand, this is an equation for this
$V(x,x')$.  On the other hand, the procedure of this section is
directly applicable to the Schr\"odinger equation, which differs from
Eq.\ (\ref{eq:5.1}) only in the absence of the $\delta(x-x')$ term on the
right-hand side. 

This section serves two distinct purposes.  First, the parameters in
the known resolvent of Sec.\ \ref{sec:3}, especially Eqs.\
(\ref{eq:3.14}) and (\ref{eq:3.16}), are to be related to the coupling
constants $g_1$, $\,g_2$, and $g_3$ in Eq.\ (\ref{eq:4.42}).  This will
give an explicit verification of consistency of the prescriptions given
in Sec.\ \ref{sec:4}.  Secondly, the procedure to be followed here
serves as a useful introduction to the slightly more complicated problem
of the next section, where a two-channel scattering by a Fermi
pseudo-potential is taken to be a model for a quantum memory.

The solution $R_{\kappa}(x,x')$ is given, as in the general case, by Eq.\
(\ref{eq:2.7}).  The substitution of Eq.\ (\ref{eq:2.7})
into Eq.\ (\ref{eq:5.1}) gives
\begin{equation}
\left(-\frac{d^2}{dx^2}+\kappa^2\right)\left[-\frac{1}{2\kappa}
\,f(\kappa;\mbox{sg}\,x,\mbox{sg}\,x')e^{-\kappa(|x|+|x'|)}\right]
+\int_{-\infty}^{\infty} dx''\,V(x,x'')R_{\kappa}(x'',x')=0.\label{eq:5.2}
\end{equation}
Since the first term has been evaluated by Eq.\ (\ref{eq:4.22}), Eq.\
(\ref{eq:5.2}) can be written alternatively as
\begin{eqnarray}
\lefteqn{2\kappa\int_{-\infty}^{\infty}
dx''\,V(x,x'')R_{\kappa}(x'',x')}\quad\nonumber\\[2pt]
&=&e^{-\kappa|x'|}\begin{cases}
\left\{-[f_1(\kappa)-f_2(\kappa)]\delta'(x)+\kappa[f_1(\kappa)
+f_2(\kappa)]\delta(x)\right\},&\mbox{for\ }x'>0\\
\left\{-[f_4(\kappa)-f_3(\kappa)]\delta'(x)+\kappa[f_4(\kappa)
+f_3(\kappa)]\delta(x)\right\},&\mbox{for\
}x'<0.\end{cases}\qquad\label{eq:5.3}
\end{eqnarray}
Using the knowledge gained from Sec.\ \ref{sec:4}, a fairly lengthy
calculation gives more explicitly
\begin{eqnarray}
\lefteqn{[g_1\delta(x)+g_2\delta'(x)]e^{-\kappa|x'|}
\begin{cases}
\left\{1-\frac{1}{2} [f_1(\kappa)+f_2(\kappa)]\right\},&\mbox{for\ }x'>0\\
\left\{1-\frac{1}{2}\,[f_4(\kappa)+f_3(\kappa)]\right\},&\mbox{for\
}x'<0\end{cases}}\quad\nonumber\\[4pt]
\lefteqn{\mbox{}+[g_2\delta(x)+g_3\delta'(x)]\kappa 
e^{-\kappa|x'|}\begin{cases}
\left\{-1+\frac{1}{2}\,[-f_1(\kappa)+f_2(\kappa)]\right\},&\mbox{for\
}x'>0\\
\left\{1+\frac{1}{2}\,[-f_4(\kappa)+f_3(\kappa)]\right\},&\mbox{for\
}x'<0\end{cases}}\quad\nonumber\\[4pt]
&=&e^{-\kappa|x'|}\begin{cases}
\left\{-[f_1(\kappa)-f_2(\kappa)]\delta'(x)+\kappa[f_1(\kappa)+f_2(\kappa)]
\delta(x)\right\},&\mbox{for\ }x'>0\\
\left\{-[f_4(\kappa)-f_3(\kappa)]\delta'(x)+\kappa[f_4(\kappa)+f_3(\kappa)]
\delta(x)\right\},&\mbox{for\ }x'<0,\end{cases}\qquad\label{eq:5.4}
\end{eqnarray} 
where Eq.\ (\ref{eq:4.45}) has been used.  In Eq.\ (\ref{eq:5.4}), all
dependences on $x'$ cancel out.  It therefore only remains to identify
the coefficients of $\delta(x)$ and $\delta'(x)$.  The results are 
\begin{subequations}
\label{eq:5.5}
\begin{eqnarray}
\lefteqn{g_1\left\{1-{\textstyle
\frac{1}{2}}\,[f_1(\kappa)+f_2(\kappa)]\right\} +\kappa
g_2\left\{-1+{\textstyle \frac{1}{2}}
[-f_1(\kappa)+f_2(\kappa)]\right\}}\quad\nonumber\\[4pt]
&=&\kappa[f_1(\kappa)
+f_2(\kappa)],\hskip2.5in\label{eq:5.5a}\\[8pt] 
\lefteqn{g_2\left\{1-{\textstyle
\frac{1}{2}}\,[f_1(\kappa)+f_2(\kappa)]\right\}+\kappa
g_3\left\{-1+{\textstyle \frac{1}{2}}
[-f_1(\kappa)+f_2(\kappa)]\right\}}\quad\nonumber\\[4pt]
&=& -[f_1(\kappa)
-f_2(\kappa)],\label{eq:5.5b}\\[8pt] 
\lefteqn{g_1\left\{1-{\textstyle
\frac{1}{2}}\,[f_4(\kappa)+f_3(\kappa)]\right\}+\kappa g_2\left\{
1+{\textstyle \frac{1}{2}}
[-f_4(\kappa)+f_3(\kappa)]\right\}}\quad\nonumber\\[4pt]
&=&
\kappa[f_4(\kappa) +f_3(\kappa)],\label{eq:5.5c}\\[8pt]
\lefteqn{g_2\left\{1-{\textstyle \frac{1}{2}}
[f_4(\kappa)+f_3(\kappa)]\right\}+\kappa g_3\left\{1 +{\textstyle
\frac{1}{2}} [-f_4(\kappa)+f_3(\kappa)]\right\}}\quad\nonumber\\[4pt]
&=&-[f_4(\kappa) -f_3(\kappa)].\label{eq:5.5d} 
\end{eqnarray}
\end{subequations} 
Solving Eqs.\ (\ref{eq:5.5}) gives
\begin{subequations}
\label{eq:5.6}
\begin{eqnarray}
f_1(\kappa)&=&D^{-1}[-g_3 \kappa +2g_2
-g_1\kappa^{-1}],\label{eq:5.6a}\\[6pt]
f_3(\kappa)&=&D^{-1}[-g_3\kappa
-2g_2-g_1\kappa^{-1}],\label{eq:5.6b}\\[6pt]
f_2(\kappa)&=&f_4(\kappa)=1+D^{-1}\,\frac{1}{2}\,(4+g_1g_3-g_2^2),
\label{eq:5.6c}
\end{eqnarray}
\end{subequations} 
where
\begin{equation}
D=g_3\kappa-\frac{1}{2}\,(4-g_1g_3+g_2^2)-g_1\kappa^{-1}.\label{eq:5.7}
\end{equation}  

Equations (\ref{eq:5.6}) are to be compared with Eqs.\ (\ref{eq:3.14})
and (\ref{eq:3.16}).  First, this gives a deeper understanding why there
are the two distinct cases (\ref{eq:3.14}) and (\ref{eq:3.16}). 
Equations (\ref{eq:3.14}) correspond to the situation where $g_1$ and
$g_3$ have the same sign, while Eqs.\ (\ref{eq:3.16}) correspond to $g_1$
and $g_3$ having opposite signs.  Secondly, in both cases, it is seen
immediately from Eqs.\ (\ref{eq:5.6}) that $c_2=c_4$, a fact that has been
used before.  The results of expressing the five $c$'s in terms of the
three
$g$'s are the following:
\begin{equation}
c_0=\sqrt{\left|{g_3/g_1}\right|},\hskip1in\label{eq:5.8}
\end{equation}
\vspace*{-18pt}
\begin{subequations} 
\label{eq:5.9} 
\begin{eqnarray}
c_1&\!\!=\!\!&\frac{1}{4}\,(4-g_1g_3+g_2^2),\label{eq:5.9a}\\[4pt]
c_2&\!\!=\!\!&c_4=\frac{1}{4}\,(4+g_1g_3-g_2^2),\label{eq:5.9b}\\[4pt]
c_3&\!\!=\!\!&g_2.\label{eq:5.9c}
\end{eqnarray}
\end{subequations}
Here use has been made of the scale invariance (\ref{eq:3.18}). 
[Strictly speaking, the right-hand sides of the three Eqs.\
(\ref{eq:5.9a}--c) should all be multiplied by the factor
$\mbox{sg}\,g_3$.  This factor has been omitted because it has no
consequences.]

For completeness and also for later use, let the scattering matrix be
written down.  This involves returning to the more familiar variable $k$
through Eq.\ (\ref{eq:2.2}) and then letting $x'\to \pm \infty$.  After
analytic continuation to positive values of $k$, the $S$-matrix is a
$2\times 2$ matrix
\begin{equation}
S=\left[\,\begin{matrix}
S_{++}\quad&S_{+-}\\[3pt]
S_{-+}\quad&S_{--}\end{matrix}\,\right],\label{eq:5.10}
\end{equation} 
where $+$ denotes propagation in the $+x$ direction, and $-$ in the $-x$
direction.  For any interaction at the point $x=0$, it follows from
Eqs.\ (\ref{eq:2.7}) and (\ref{eq:2.8}) that
\begin{eqnarray}
S&=&\left[\,\begin{matrix} 
1-f(-ik;+,-)\quad&-f(-ik;-,-)\\[3pt]
-f(-ik;+,+)\quad&1-f(-ik;-,+)\end{matrix}\,\right]\nonumber\\[4pt]
& =&
\left[\,\begin{matrix}
1-f_4(-ik)\quad&-f_3(-ik)\\[3pt]
-f_1(-ik)\quad&1-f_2(-ik)\end{matrix}\,\right].\label{eq:5.11}
\end{eqnarray}
Equations (\ref{eq:5.6}) and (\ref{eq:5.7}) then give explicitly
\begin{eqnarray}
S&=&\left[ig_3k+\frac{1}{2}\,(4-g_1g_3+g_2^2)+ig_1k^{-1}
\right]^{-1}\nonumber\\[4pt]
&&\mbox{}\times\left[\,\begin{matrix}
\frac{1}{2}\,(4+g_1g_3-g_2^2)\quad&ig_3k-2g_2-ig_1k^{-1}\\[3pt]
ig_3k+2g_2-ig_1k^{-1}\quad&\frac{1}{2}\,(4+g_1g_3-g_2^2)\end{matrix}
\,\right],\label{eq:5.12}
\end{eqnarray}
which is unitary.

An interesting special case is that with $g_2=0$; in this case, the
pseudo-potential is even and there is left-right symmetry.  Explicitly,
in this case $g_2=0$, the $S$-matrix is 
\begin{equation}
S=\left[ig_3k+\frac{1}{2}\,(4-g_1g_3)+ig_1k^{-1}\right]^{-1}
\left[\,\begin{matrix}
\frac{1}{2}\,(4+g_1g_3)\quad&ig_3k-ig_1k^{-1}\\[3pt]
ig_3k-ig_1k^{-1}\quad&\frac{1}{2}\,(4+g_1g_3)\end{matrix}
\,\right].\label{eq:5.13}
\end{equation}
This special case $g_2=0$, generalized to the case of coupled channels,
is going to play a central role in Part \ref{part:B} of this paper.

This completes the present discussion of the theory of Fermi
pseudo-potential in one dimension.  Attention is now turned to the
first application of this theory. 

\part{\large\textbf{Application of One-Dimensional Fermi
Pseudo-Potential}}\label{part:B}

\section{Model for Quantum Memory}\label{sec:6}

There are many possible applications of the Fermi pseudo-potential in one
dimension.  As an example, it is intriguing to ask under what conditions,
if any, Bethe's hypothesis \cite{bethe1931,yangyang1969} still holds when
the delta-function potential is replaced by the $V(x,x')$ of Eq.\
(\ref{eq:4.42}).  As a first application, however, it is more desirable
to begin with a case where the Fermi pseudo-potential is used in a
relatively simple situation of current interest.

For decades, computer components have become smaller and smaller, and this
trend is expected to continue \cite{moore1965}.  When some of the
components become sufficiently small, as to be expected in the
not-too-distant future, they need to be described in general by quantum
mechanics.  No matter how quantum computing is to develop in the future,
one important component is necessarily the quantum memory, sometimes
called the quantum register.  The main function of any quantum memory is
to store a quantum state.

In order for a quantum memory to be useful, it must be possible to alter
the quantum state in the memory in a controlled way.  This can only be
accomplished by sending a signal from outside of the memory.  In other
words, the quantum state in the memory is to be controlled by a scattering
process \cite{pikesabatier2002}.

It is the purpose of Sec.\ \ref{sec:6} to propose a simple model for
quantum memory.  First, in order to have scattering processes, at least
one space dimension is necessary.  Otherwise there is no possibility of
interference between the incident wave and the scattered wave.  As
perhaps to be expected, this interference is of central importance. 
Since the state in the memory must include at least two independent
quantum states, it is simplest to describe the quantum memory using the
coupled Schr\"odinger equations for two channels.  This is essentially
Eq.\ (\ref{eq:1.1}) in the Introduction.

It remains to make the simplest choice for the $2\times 2$ matrix
potential $V(x)$ of Eq.\ (\ref{eq:1.2}).  This simplest choice, the Fermi
pseudo-potential in one dimension, has been investigated systematically
in Part~\ref{part:A} of this paper, the general result being given by
Eq.\ (\ref{eq:4.42}).  

The symmetry properties of this $V(x,x')$ under space inversion have been
discussed at the end of Sec.~\ref{sec:4}.  In particular, it is
symmetrical if $g_2=0$.  In order for the model to be suitable for
quantum memory, it is essential to concentrate on this special case.  The
reason is that, only in this case, do the even wave $\cos kx$ and the odd
wave $\sin kx$ not mix.  This is also the basis for the comment in the
Introduction, after Eq.\ (\ref{eq:1.3}), why the $\delta'(x)$ potential
is not suitable for quantum memory.  That this absence of mixing is
important is discussed further in Sec.\ \ref{sec:7} for a more general
setting.

With this understanding and choice, the present model for the quantum
memory is described by the one-dimensional coupled Schr\"odinger
equations
\begin{eqnarray}
-\frac{d^2\psi_1(x)}{dx^2}+\int_{-\infty}^{\infty}dx'\,[V_{11}(x,x')
\psi_1(x') +V_{12}(x,x')\psi_2(x')]&=&k^2\psi_1(x),\nonumber\\[8pt]
-\frac{d^2\psi_2(x)}{dx^2}+\int_{-\infty}^{\infty}dx'\,[V_{21}(x,x')
\psi_1(x') +V_{22}(x,x')\psi_2(x')]&=&k^2\psi_2(x),\label{eq:6.1}
\end{eqnarray}
with the $2\times 2$ matrix potential 
\begin{eqnarray}
V(x,x')&=&\left[\,\begin{matrix}
V_{11}(x,x')\quad&V_{12}(x,x')\\[4pt]
V_{21}(x,x')\quad&V_{22}(x,x')\end{matrix}
\,\right]\nonumber\\[8pt]
&=&\left[\,\begin{matrix}
g_3\delta'(x)\delta'(x')\quad&g_1\delta(x)\delta(x')\\[4pt]
g_1\delta(x)\delta(x')\quad&-g_3\delta'(x)\delta'(x')\end{matrix}
\,\right].\label{eq:6.2}
\end{eqnarray}
A more elegant way to write this potential is
\begin{equation}
V(x,x')=g_1\delta(x)\delta(x')\sigma_1+g_3\delta'(x)\delta'(x')\sigma_3,
\label{eq:6.3}
\end{equation}
where the $\sigma$'s are the Pauli matrices.

When $g_2=0$, the potential $g_1\delta(x)\delta(x')$ does not act on the
odd wave, and similarly the potential $g_3\delta'(x)\delta'(x')$ does not
act on the even wave.  The first part of this claim is easy to obtain, and
the second part follows from the definition (\ref{eq:4.28}) of
$\delta'_p(x)$.  Alternatively, they can be seen from Eq.\
(\ref{eq:5.13}), where $S_{++}=S_{--}$ and $S_{+-}=S_{-+}$.  For the even
wave, the scattering phase shift is given by
\begin{equation}
S_{++}+S_{+-}=\frac{2k-ig_1}{2k+ig_1}\label{eq:6.4}
\end{equation}
independent of $g_3$, while, for the odd wave, it is
\begin{equation}
S_{++}-S_{+-}=\frac{2-ig_3k}{2+ig_3k}\label{eq:6.5}
\end{equation}
independent of $g_1$.  Therefore, for the present case of two coupled
channels as described by Eqs.\ (6.1) and (6.2), the $S$-matrix for the
even and odd cases can be expressed in terms of these quantities as
follows.  Consider first the case for the odd wave; since the $g_1$ term
does not contribute and can be neglected, the $V(x,x')$ of Eq.\
(\ref{eq:6.2}) effectively reduces to
\begin{equation}
V(x,x')=\left[\,\begin{matrix}
g_3\delta'(x)\delta'(x')\quad&0\\[4pt]
0\quad&-g_3\delta'(x)\delta'(x')\end{matrix}
\,\right],\label{eq:6.6}
\end{equation}
which is diagonal, meaning that $\psi_1(x)$ and $\psi_2(x)$ do not
couple.  Since the behaviors of the two channels differ only in the sign
of $g_3$, the $S$-matrix for this odd case is given by Eq.\
(\ref{eq:6.5}), or more explicitly
\begin{equation}
S_-(k)=\left[\,\begin{matrix}
{\displaystyle \frac{2-ig_3k}{2+ig_3k}}\quad&0\\[6pt]
0\quad&{\displaystyle \frac{2+ig_3k}{2-ig_3k}}\end{matrix}
\,\right].\label{eq:6.7}
\end{equation}
It is instructive to rewrite this expression in terms of $\sigma_3$:
\begin{equation}
S_{-}(k)=\frac{(4-g_3^2k^2)-4ig_3k\sigma_3}{4+g_3^2k^2}=\exp\left[-i
\sigma_3
\left(2\tan^{-1}\,\frac{g_3k}{2}\right)\right].\label{eq:6.8}
\end{equation}
For the even wave, it is merely necessary to replace the right-hand
side of Eq.\ (\ref{eq:6.5}) by that of Eq.\ (\ref{eq:6.4}), and also
$\sigma_3$ by $\sigma_1$.  Therefore Eq.\ (\ref{eq:6.8}) leads to
\begin{equation}
S_{+}(k)=\frac{(4k^2-g_1^2)-4ig_1k\sigma_1}{4k^2+g_1^2}=\exp\left[-i
\sigma_1\left( 2\tan^{-1}\,\frac{g_1}{2k}\right)
\right]\label{eq:6.9}
\end{equation}
or 
\begin{equation}
S_{+}(k)=\left[\,\begin{matrix}
{\displaystyle \frac{4k^2-g_1^2}{4k^2+g_1^2}}\quad&{\displaystyle
\frac{-4ig_1k}{4k^2+g_1^2}}\\[12pt]
{\displaystyle \frac{-4ig_1k}{4k^2+g_1^2}}\quad&{\displaystyle
\frac{4k^2-g_1^2}{4k^2+g_1^2}}\end{matrix}
\,\right].\label{eq:6.10}
\end{equation}   
When neither $g_1$ nor $g_3$ is zero, any given element $S$ of SU(2) can
be expressed as a finite product of $S_{+}(k)$ and $S_{-}(k)$, i.e.,
\begin{equation}
S=S(k_1)\,S(k_2)\,\cdots\,S(k_m),\label{eq:6.11}
\end{equation}
where each $S(k_i)$ is suitably chosen as $S_{+}(k_i)$ or $S_{-}(k_i)$. 

In the language of scattering theory, the meaning of $S_{+}(k)$ is as
follows.  [The meaning of $S_{-}(k)$ is similar.]  The ``in'' field is
\begin{equation}
\Psi^{\text{in}}=\left[\,\begin{matrix}
a_1^{\text{in}}\\[4pt]
a_2^{\text{in}}\end{matrix}\,\right] e^{-ik|x|},\label{eq:6.12}
\end{equation}
while the ``out'' field is
\begin{equation}
\Psi^{\text{out}}=\left[\,\begin{matrix}
a_1^{\text{out}}\\[4pt]
a_2^{\text{out}}\end{matrix}\,\right] e^{ik|x|}.\label{eq:6.13}
\end{equation}
Then
\begin{equation}
\left[\,\begin{matrix} a_1^{\text{out}}\\[4pt] a_2^{\text{out}}\end{matrix}
\,\right]=S_{+}(k)\left[\,\begin{matrix}
a_1^{\text{in}}\\[4pt] a_2^{\text{in}}\end{matrix}\,\right].\label{eq:6.14}
\end{equation} 
In other words, the quantum state in the memory is
$\Bigl[\,{\begin{matrix} a_1^{\text{in}}\\ a_2^{\text{in}}\end{matrix}}
\,\Bigr]$ before scattering,\vadjust{\kern-8pt} and $\Bigl[\,{\begin{matrix} a_1^{\text{out}}\\ a_2^{\text{out}}\end{matrix}}
\,\Bigr]$ after scattering.  These states before and after scattering
are related by Eq.\ (\ref{eq:6.14}).  

For any memory, classical or quantum, the basic operations are
\textit{write}, \textit{read}, and \textit{reset}.  Of these three
operations, writing is the simplest:  Given the initial state
$\Bigl[\,{\begin{matrix} a_1^{\text{in}}\\
a_2^{\text{in}}\end{matrix}}\,\Bigr]$ and the desired final state
$\Bigl[\,{\begin{matrix} a_1^{\text{out}}\\
a_2^{\text{out}}\end{matrix}}\,\Bigr]$, there is a desired $S$-matrix
$S$.  Express this particular $S$ by Eq.\ (\ref{eq:6.11}) as a finite
product; then the writing is accomplished by a sequence of these $m$
scatterings.

The question may be asked:  While the final state
$\Bigl[\,{\begin{matrix} a_1^{\text{out}}\\
a_2^{\text{out}}\end{matrix}}\,\Bigr]$ is the desired\vadjust{\kern-8pt}
state to be stored in the quantum memory, how can one know what the
initial state $\Bigl[\,{\begin{matrix} a_1^{\text{in}}\\ a_2^{\text{in}}
\end{matrix}}\,\Bigr]$ is?  This is where the idea of a standard state
$s$ is needed.  ``Resetting'' means changing the content of the quantum
memory, whatever it is, to the standard state $s$.  For writing, the
initial state is the standard state, i.e., 
\begin{equation}
\left[\,\begin{matrix}
a_1^{\text{in}}\\[4pt]
a_2^{\text{in}}\end{matrix}
\,\right]=s.\label{eq:6.15}
\end{equation} 
In other words, before writing on a quantum memory, it is first reset so
that Eq.\ (\ref{eq:6.15}) is satisfied.  The standard state can be chosen
to be any quantum state; however, once chosen, the choice is rarely
altered.

Since scattering from a quantum memory leads to a unitary transformation
of the quantum state in the memory, resetting cannot be accomplished
without first finding out the content of the quantum memory.  In other
words, the first step of ``resetting'' is ``reading.''  After the content
of the quantum memory is known, say $\Bigl[\,{\begin{matrix}
a_1^{\text{in}}\\ a_2^{\text{in}}\end{matrix}}\,\Bigr]$, ``resetting''
involves finding a sequence of scattering $S(k_j)$, $\,j=1$, 2, \dots\
$n$, via Eq.\ (\ref{eq:6.11}) such that the resulting $S$ has the property
\begin{equation}
S\left[\,\begin{matrix} a_1^{\text{in}}\\[4pt] a_2^{\text{in}}\end{matrix}
\,\right]=s.\label{eq:6.16}
\end{equation}
In summary, if ``reading'' can be accomplished, then so can
``resetting''; if ``resetting'' can be accomplished, so can ``writing.''

The main task here is therefore to discuss, within the present model, the
operation of reading a quantum memory.  More precisely, what is involved
is the following.  When the quantum state in a memory,
$\Bigl[\,{\begin{matrix} a_1^{\text{in}}\\
a_2^{\text{in}}\end{matrix}}\,\Bigr]$, is not known, find a suitably
chosen sequence of incident waves $\cos k_ix$ or $\sin k_ix$ such that
the knowledge about the field can be used to determine the values of
$a_1^{\text{in}}$ and
$a_2^{\text{in}}$.  After this determination, the quantum memory is
returned to the initial state
$\Bigl[\,{\begin{matrix} a_1^{\text{in}}\\ a_2^{\text{in}}\end{matrix}}
\,\Bigr]$.  [The last step is similar to the classical case in which a
core memory is read from an initial state and then returned to it.]

Let the quantum state in the memory be $\Bigl[\,{\begin{matrix} a_1\\
a_2\end{matrix}}\,\Bigr]$; the problem is to determine the values of
$a_1$ and $a_2$ by scattering from this state.  Suppose an odd wave is
used for the first scattering; then the two-component wave function for
$x>0$ is given explicitly by 
\begin{equation}
\psi(x)=\left[\,\begin{matrix} \psi_1(x)\\[4pt] \psi_2(x)\end{matrix}
\,\right] =\left[\,\begin{matrix} a_1\\[4pt] a_2\end{matrix}\,\right]
e^{-ikx} +\left[\,\begin{matrix}a_1e^{i\phi_{-}}\\[4pt]
a_2e^{-i\phi_{-}}\end{matrix}
\,\right]e^{ikx},\label{eq:6.17}
\end{equation}
where, by Eq.\ (\ref{eq:6.8}), 
\begin{equation}
\phi_{-}=-2\tan^{-1}\,\frac{1}{2}\,g_3k.\label{eq:6.18}
\end{equation}
In particular,
\begin{equation}\psi(x)^{\dag}\psi(x)=2[1+\cos
2kx\cos\phi_{-}-(|a_1|^2-|a_2|^2)\sin 2kx \sin \phi_{-}],\label{eq:6.19}
\end{equation}
because $|a_1|^2 +|a_2|^2=1$.  Therefore, this scattering process gives,
through the interference term, the quantity 
\begin{equation}
A_1=|a_1|^2-|a_2|^2.\label{eq:6.20}
\end{equation}
Similarly, if the quantum state in the memory is first returned to the
original state by a suitable scattering, then a second scattering with an
even wave gives, again through the interference term, the second quantity
\begin{equation}
A_2=\frac{1}{2}\,[|a_1+a_2|^2 +|a_1-a_2|^2]=2\
\mbox{Re}\,a_1^*a_2.\label{eq:6.21}
\end{equation}
These two quantities, $A_1$ and $A_2$, are sufficient to determine the
values of the complex numbers $a_1$ and $a_2$, except for a common
phase.  In order to determine this common phase, it is simplest to use
the known standard state $s$.  For example, a further interference with
this standard state using, say, the odd wave gives a third quantity
\begin{equation}
A_3=|a_1+s_1|^2-|a_2+s_2|^2.\label{eq:6.22}
\end{equation}
Since $s_1$ and $s_2$ are known, these three quantities $A_1$, $\,A_2$
and $A_3$ determine $a_1$ and $a_2$.  Returning once more to the original
quantum state presents no problem.

This completes the description of the present model of the quantum
memory, including the operations of writing, reading, and resetting.

The advantages of this model, based on the Fermi pseudo-potential in one
dimension, are its simplicity and its being completely explicit.  On the
one hand, such an explicit model plays an essential role in the initial
understanding of some aspects of a new problem.  On the other hand, the
usefulness of such a model really lies in the possibility of opening a
line of inquiry into these aspects.  This is to be discussed in some
detail in the next section.  That is, in Sec.\ \ref{sec:7} an attempt is to
be made to present a general picture concerning the quantum memory,
emphasizing the operations of writing, reading, and resetting, all
accomplished by repeated scattering.

Some simplifying assumptions introduced in the model of this section are
clearly not needed in the general setting of the next section.  An
example is the choice of using the Fermi pseudo-potential $V(x,x')$ of
Eq.\ (\ref{eq:6.2}); another one is the use of the Schr\"odinger equation
(\ref{eq:6.1}) in one dimension.  Thus the generalization to the
Schr\"odinger equation in three dimensions with a more general potential
is immediate but the results are less explicit.  The further
generalization to renormalized quantum field theory also does not
present any obstacle.

What is less clear, and most important, is the role played by the
condition $g_2=0$, used throughout this section.  This condition is
closely related to, and makes it possible to use, the even waves and
the odd waves.  In order to appreciate this point, take instead the
incoming wave as, say, from the direction of the $-x$ axis, i.e.,
\begin{equation}
\Psi^{\text{in}}=\left[\,\begin{matrix} a_1^{\text{in}}\\[4pt] 
a_2^{\text{in}}\end{matrix}\,\right]e^{-ikx}.\label{eq:6.23}
\end{equation}
This is a superposition of an even wave and an odd wave.  Since the
Schr\"odinger equation is linear, the even part is operated on by the
$S_{+}(k)$ of Eq.\ (\ref{eq:6.10}), and the odd part by the
$S_{-}(k)$ of Eq.\ (\ref{eq:6.7}).  Since $g_1$ and $g_3$ are not zero
and thus these $S_{+}(k)$ and $S_{-}(k)$ are not equal, the quantum
state in the memory for an outgoing wave in the $+x$ direction is
different from that for an outgoing wave in the $-x$ direction.  In
other words, in order to determine the quantum state in the memory after
scattering with the $\Psi^{\text{in}}$ of Eq.\ (\ref{eq:6.23}), it is
necessary to detect the direction of the outgoing wave. 

In order for a quantum memory to behave as a memory, i.e., as the
storage for a quantum state, it is essential that what is in the memory
does not depend on the behavior of the scattered wave.  Indeed, from
the point of view of scattering theory, this characterizes quantum
memories.  Therefore, for the present model with the Fermi
pseudo-potential, some incident waves, such as the even wave and the
odd wave, are acceptable or ``admissible,'' while many others, such as
$e^{-ikx}$ of Eq.\ (\ref{eq:6.23}), are not ``admissible.'' 

This concept of admissible incident waves is central, not only for the
present model but also in general.  This is the first topic to be
discussed in the next section.

\section{Generalization}\label{sec:7}

It is the purpose of this section to give a general description of
quantum memories.  This is to be accomplished by extracting the dominant
features from the model of Sec.\ \ref{sec:6} on the basis of the Fermi
pseudo-potential in one dimension.

In order to extract the dominant features, consider first the 
following two generalizations, the first one obvious and the second 
one less so.

First, that the potential is the Fermi pseudo-potential is not 
necessary.  In other words, the matrix potential $V(x, x')$ of Eq.\
(\ref{eq:6.2}) can take a fairly general form.  That $g_2$ is zero
translates into the  condition that this $V(x, x')$ is symmetrical, i.e.,
\begin{displaymath}
V(-x, -x') = V(x, x')
\end{displaymath}
in general.

Secondly, that the model is one-dimensional is not essential.  For 
example, the model can be a two-channel scattering in three-dimensional 
space.  In this case, the $V(x, x')$ is replaced by another $2\times 2$
matrix  potential $V(\textbf{r},\textbf{r}')$, while the symmetry of the $V(x,
x')$ becomes the condition that this $V(\textbf{r},\textbf{r}')$ is rotationally
symmetrical.
     
While this rotational symmetry is probably not necessary, this  symmetry
does play an important role.  In the one-dimensional case  studied in
detail in Sec.\ \ref{sec:6}, the symmetry of the $V(x, x')$, coming from 
$g_2 = 0$, makes it possible to use the even wave and the odd wave.  
Similarly, in three dimensions, the rotational symmetry of
$V(\textbf{r},\textbf{r}')$ makes it possible to use partial waves: the
various partial waves do not couple so that each partial ``in'' wave
leads to only the corresponding  partial ``out'' wave.  
     
Consider now the more general setting.  Let the quantum memory be in a
pure state $\sum_j a_j|j\rangle$, where $|j\rangle$ is a complete set of
linearly independent states for the memory.  The standard state $s$ is a
particular linear combination of these $|j\rangle$.  Let $\psi$ denote
the wave function sent in from the outside to interact with the quantum
memory;  then the ``in'' field for the scattering process on the memory is
\begin{equation}
\Psi^{\text{in}}=\left( \sum_j a_j^{\text{in}}|j\rangle\right)
\psi^{\text{in}}.\label{eq:7.1}
\end{equation}
An example of $\Psi^{\text{in}}$ is given by Eq.\ (\ref{eq:6.23}).  It
should be emphasized that $\psi^{\text{in}}$ is at our disposal to
accomplish whatever the purpose of this scattering is.

It is the fundamental characteristic of the scattering from a quantum
memory that not only is the ``in'' field $\Psi^{\text{in}}$ of the form of
Eq.\ (\ref{eq:7.1}), but also the ``out'' field is of a similar form,
\begin{equation}
\Psi^{\text{out}}=\left(\sum_j a_j^{\text{out}}|j\rangle\right)
\psi^{\text{out}}.\label{eq:7.2}
\end{equation}
For the special model of Sec.\ \ref{sec:6}, this important point has been
discussed near the end of that section.  For the present generalization,
it is worked out in detail in Appendix~\ref{app:B}.  As already seen in
Sec.\ \ref{sec:6}, Eq.\ (\ref{eq:7.2}) puts strong conditions on
$\psi^{\text{in}}$.  More precisely, a $\psi^{\text{in}}$ is defined to be
\textit{admissible} if, for all $\sum_j a_j^{\text{in}}|j\rangle$, the
corresponding $\Psi^{\text{out}}$ is a tensor product as given by Eq.\
(\ref{eq:7.2}).

It should be added parenthetically that this definition of being
admissible can be easily generalized by restricting the $\sum_j
a_j^{\text{in}}|j\rangle$ to certain subsets.  This generalization is
expected to be useful in future investigations, but is not needed for
this paper.

In order to perform the operations of writing, reading, and resetting a
quantum memory, it is necessary to have a sufficiently large collection
of admissible $\psi^{\text{in}}$.  This has been verified to be the case
for the model of Sec.\ \ref{sec:6}, and will be assumed to be so in this
section.  Let
\begin{equation}
\psi^{\text{in}}\left(\sum_j a_j^{\text{in}}|j\rangle\to \sum_j
a_j^{\text{out}}|j\rangle\right)\label{eq:7.3}
\end{equation}
denote a $\psi^{\text{in}}$ with the property that, if Eq.\ (\ref{eq:7.1})
holds for this $\psi^{\text{in}}$, then Eq.\ (\ref{eq:7.2}) holds.  It is
assumed that, given any $a_j^{\text{in}}$ and $a_j^{\text{out}}$, there
is at least one such $\psi^{\text{in}}$.  It is possible that there is
more than one such $\psi^{\text{in}}$.  It has been seen from the model
of Sec.\ \ref{sec:6} that this $\psi^{\text{in}}$ may actually involve a
sequence of $\psi^{\text{in}}$'s; see especially Eq.\ (\ref{eq:6.11}). 
However, for simplicity of notation, the expression (\ref{eq:7.3}) will
be retained.

The operations of writing, reading, and resetting are now to be described
in this order.  For the purpose of writing after the quantum memory has
been reset to the standard state $s$, it is sufficient to use any one of
the $\psi^{\text{in}}\bigl(s\to \sum_j a_j|j\rangle\bigr)$, where $\sum_j 
a_j|j\rangle$ is the desired quantum state to be put in the memory. 

Reading from a quantum memory is more complicated.  Let a quantum memory
be in a state $\sum_j a_j^{\text{in}}|j\rangle$; it is desired to determine
the values of these $a_j^{\text{in}}$ by interrogating this memory, i.e.,
by sending a suitably chosen sequence of admissible $\psi^{\text{in}}$'s, 
\begin{equation}
\psi^{\text{in}(1)},\ \psi^{\text{in}(2)},\ \psi^{\text{in}(3)},\cdots,\
\psi^{\text{in}(N)},\label{eq:7.4}
\end{equation}
and scattering them successively by this quantum memory.  More precisely,
consider the successive scattering processes  
\begin{eqnarray}
\sum_j a_j^{\text{in}}|j\rangle \psi^{\text{in}(1)}\to \sum_j 
a_j^{\text{out}(1)}|j\rangle\psi^{\text{out}(1)}\to
a_j^{\text{out}(1)}&=&a_j^{\text{in}(2)}\nonumber\\[4pt]
\to \sum_j a_j^{\text{in}(2)}|j\rangle\psi^{\text{in}(2)}\to \sum_j 
a_j^{\text{out}(2)}|j\rangle\psi^{\text{out}(2)}\to
a_j^{\text{out}(2)}&=&a_j^{\text{in}(3)}\nonumber\\[4pt]
\to \sum_j a_j^{\text{in}(3)}|j\rangle\psi^{\text{in}(3)}\to \sum_j 
a_j^{\text{out}(3)}|j\rangle\psi^{\text{out}(3)}\to 
a_j^{\text{out}(3)}&=&a_j^{\text{in}(4)}\nonumber\\[4pt]
\to\cdots\hskip3.17in\nonumber\\[4pt]
\to \sum_j a_j^{\text{in}(N)}|j\rangle\psi^{\text{in}(N)}\to \sum_j
a_j^{\text{out}(N)}|j\rangle\psi^{\text{out}(N)}.\hskip.425in\label{eq:7.5}
\end{eqnarray} 
Corresponding to the list (\ref{eq:7.4}), there is a list of
$\psi^{\text{out}}$'s, 
\begin{equation}
\psi^{\text{out}(1)},\ \psi^{\text{out}(2)},\ \psi^{\text{out}(3)},\
\cdots,\ \psi^{\text{out}(N)}.\label{eq:7.6}
\end{equation}
From the quantities given in (\ref{eq:7.4}) and (\ref{eq:7.6}) together
with their interference, the values of $a_j^{\text{in}}$ are obtained. 
This has been demonstrated explicitly in Sec.\ \ref{sec:6} for the model
there, and it is also shown there that a further interference with the
standard state $s$ may be needed to determine the overall phase.  The
importance of interference cannot be over-emphasized.

Once the $a_j^{\text{in}}$ are known, the values of $a_j^{\text{out}(N)}$
of the process (\ref{eq:7.5}) can be obtained.  An additional scattering
using any one of the admissible $\psi^{\text{in}}\bigl(\sum_j
a_j^{\text{out}(N)}|j\rangle\to \sum_ja_j^{\text{in}}|j\rangle\bigr)$
returns the quantum memory to its initial state. 

With the above process of reading a quantum memory, resetting is now
straightforward.  Resetting a quantum memory in the initial state
$\sum_j a_j^{\text{in}}|j\rangle$ to the standard state $s$ consists of
the following two steps.

(i) Read the memory to determine $a_j^{\text{in}}$.  Note that, after the
process of reading is performed, the memory is in the original initial
state $\sum_j a_j^{\text{in}}|j\rangle$.

(ii) Apply an additional scattering using any one of the admissible
$\psi^{\text{in}}\bigl(\sum_j a_j^{\text{in}}|j\rangle\to s\bigr)$.

This completes the description of the quantum memory together with
writing, reading, and resetting, all performed through scattering from
the memory.

It may be worthwhile to emphasize that the concept of the quantum memory
introduced and described here is quite general.  In particular, the
scattering process
\begin{equation}
\Psi^{\text{in}}=\left(\sum_j a_j^{\text{in}}|j\rangle\right)
\psi^{\text{in}}\to
\Psi^{\text{out}}=\left(\sum_j a_j^{\text{out}}|j\rangle\right)
\psi^{\text{out}}\label{eq:7.7}
\end{equation}
does not have many restrictions, and may or may not be linear.  Also, the
linearly independent states $|j\rangle$ are allowed to depend on time,
and may or may not be the eigenstates of an operator.

\section{Comparison with an Earlier Model}\label{sec:8}

The idea of quantum computing was first discussed by Benioff in
1980 \cite{benioff1980}.  In this pioneering paper, spatial dependence
was retained, although not quite in the form of the Schr\"odinger
equation.  Since then, quantum computing and quantum information have
become popular subjects with a vast literature \cite{LongList}. 
However, in the majority of the theoretical papers on quantum computing,
spatial dependence is omitted entirely.  Therefore the usual model for
quantum memory consists of a spin system or its generalization, and the
operations on the quantum memory consist of applying unitary matrices. 
This prevailing model for the quantum memory has led to a number of
important results.

In the present paper, as a first application of the Fermi
pseudo-potential in one dimension, an alternative model for the quantum
memory is proposed.  This model differs from the previous one mainly in
the re-introduction of the spatial variables, much in the spirit of the
original work of Benioff \cite{benioff1980}.  From the point of view of
physics, the spatial variables are clearly present, whether one wants
them or not.  Instead of saying that a unitary matrix is applied
mathematically to the content of the quantum memory, here the content of
the quantum memory is altered in a controlled way by applying suitably
chosen scatterings to the memory.

This is much more than a change of language.  While the previous model
has the advantage of simplicity, which is important because quantum
computing is a difficult subject, the present model with the spatial
variable or variables may be considered to be desirable from the
following two points of view.  First, it offers a closer description of
the experimental situation.  Since a quantum memory is necessarily small
in size, for practical reasons scattering is the simplest means of
modifying the content of a quantum memory.  Secondly, the presence of the
spatial dimensions allows more possibilities of analyzing the quantum
memory.  It is also worth mentioning that the theory of scattering has
been developed over many decades and is well understood, in the context
of both quantum mechanics and quantum field theory.  It is often
advantageous to be able to make use of existing knowledge to study a new
subject.

In both the previous model and the present model, the content of a
quantum memory is given as a pure state.  This content is altered by
applying a unitary transformation, directly in the previous model and
indirectly through scattering in the present model.  The incident,
scattered, and total wave functions have no analog in the previous
model.  In general, the phase shift \cite{blattweisskopf} of scattering is
determined from the total wave function, and the analysis of the explicit
model in Sec.\ \ref{sec:6} is actually an especially simple application
of the usual phase-shift analysis, including the prominent role played by
interference.  The point is that, while in the definition of an
admissible $\psi^{\text{in}}$ in Sec.\ \ref{sec:7} both $\Psi^{\text{in}}$
[Eq.\ (\ref{eq:7.1})] and $\Psi^{\text{out}}$ [Eq.\ (\ref{eq:7.2})] are
unentangled so far as the memory and the interrogating wave are concerned,
this is not true of the total wave function, which is for example
\begin{displaymath}
\psi(x)=\left[\,\begin{matrix}
\psi_1(x)\\
\psi_2(x)\end{matrix}
\,\right]
\end{displaymath}
for the model of Sec.~\ref{sec:6}.

There are many interesting open questions for the present model.  The
analysis of these questions is beyond the scope of the present paper. 
Nevertheless, here are two examples of such open questions.

(a) In Sec.\ \ref{sec:7}, it is explicitly assumed that there is a
sufficiently large class of admissible $\psi^{\text{in}}$ of the form
(\ref{eq:7.3}).  In the model of Sec.\ \ref{sec:6}, such a large class
indeed exists in the form of even waves and odd waves.  On the other
hand, when $g_2\ne 0$, no such large class exists.  What is needed is a
more general discussion as to the conditions under which such a
sufficiently large class of admissible $\psi^{\text{in}}$ is actually
available.

Even though examples where such a large class is available are known both
in one dimension and in three dimensions, the three-dimensional case seems
rather difficult to achieve experimentally.  If this observation is true
in general, then there may well be significant advantages to connecting
the various components of a quantum computer, including quantum memories,
by single-mode optical fibers.  In particular, sending signals through
space rather than fibers may lead to unexpected problems. 

(b) Another especially challenging and interesting question for the
present model of quantum memory concerns the issue of the so-called
no-cloning theorem.  This has been derived in the context of the previous
model, but such derivations do not seem to be applicable directly to the
present model.  This is again related to the fact that here there is not
only an $S$-matrix but also the incident, scattered, and total wave
functions.

Preliminary analysis indicates that whether the no-cloning theorem holds
for the present model of quantum memory may depend on subtle aspects of
the Schr\"odinger equation.  If this is indeed the case, then the
no-cloning theorem may need to be stated properly and precisely before it
can be derived within the present model of quantum memory.

\section{Discussions}\label{sec:9}

 The present investigation began as an attempt to understand the
$\delta'(x)$ potential in the context of the one-dimensional
Schr\"odinger equation.  When simple attempts failed, the powerful method
of the resolvent equation was used. The surprise is that, not only can the
resolvent equation be solved in general in terms of rational functions,
but also the solution yields, in addition to the well-known
$\delta$-function potential, not one but \textit{two} linearly independent
Fermi pseudo-potentials in one dimension.  One of the pseudo-potentials
is odd under space reflection and is the proper interpretation of the
$\delta'(x)$ potential.  The other one is originally unexpected and is
even under space reflection.

It is likely that there are many applications of these pseudo-potentials
to one-dimensional problems.  A possible use in statistical mechanics
connected with the Bethe ansatz \cite{bethe1931} has already
been mentioned in Sec.~\ref{sec:6}.  In this paper, only the simplest
application is discussed.  This has nothing to do with the proper
interpretation of the
$\delta'(x)$ potential, but depends critically on the unexpected, even
pseudo-potential.  By combining this even pseudo-potential with the
$\delta$-function potential, an elegant special case is found for the
scattering in two coupled channels.  Even though the two channels cannot
be decoupled, it is easy to write down the complete solution from the
known one-channel case.

In spite of the mathematical simplicity of this application of the Fermi
pseudo-potential in one dimension, this example gives a model for the
quantum memory (sometimes called the quantum register).  While this model
is completely explicit, its more important function is to point out a way
to gain a general picture concerning the quantum memory.

More generally, the time-independent Schr\"odinger equation for $n$
coupled channels with interaction at only the one point $x=0$ is
\begin{equation}
-\frac{d^2\psi(x)}{dx^2}+\int_{-\infty}^{\infty}
dx'\,V(x,x')\psi(x')=k^2\psi(x)\label{eq:9.1}
\end{equation}
with
\begin{equation}
\psi(x)=\left[\,\begin{matrix}
\psi_1(x)\\[4pt]
\psi_2(x)\\[4pt]
\vdots\\[4pt]
\psi_n(x)\end{matrix}
\,\right]\label{eq:9.2}
\end{equation}
and
\begin{equation}
V(x,x')=C_1\delta(x)\delta(x')
+C_2[\delta'_p(x)\delta(x')+\delta(x)\delta'_p(x')] +C_3\delta'_p(x)
\delta'_p(x').\label{eq:9.3}
\end{equation}
Here $C_1$, $\,C_2$ and $C_3$ are three numerical hermitian $n\times n$
matrices, while $\delta'_p(x)$ is similar to $\delta'(x)$ and is defined
in Sec.\ \ref{sec:4}.  For a given $k$ and a given incident wave
$\psi_0(x)$ with $n$ components, the solution of Eq.\ (\ref{eq:9.1})
takes the form, for $j=1$, 2, \dots, $n$,
\begin{equation}
\psi_j(x)=\psi_{0j}(x) +\begin{cases}
F_{j+}e^{ikx},\quad&\mbox{for\ }x>0\\[3pt]
F_{j-}e^{-ikx},\quad&\mbox{for\ }x<0,\end{cases}\label{eq:9.4} 
\end{equation}
analogous to Eq.\ (\ref{eq:2.7}), where the $F$'s are $2n$ coefficients
that depend on $\psi_{0j}$ and $k$.  The substitution of Eq.\
(\ref{eq:9.4}) into Eq.\ (\ref{eq:9.1}) shows that these $2n$ $F$'s
satisfy $2n$ linear equations.  Indeed, it is the power of the Fermi
pseudo-potential that Schr\"odinger equations reduce to linear algebraic
equations.  It will be interesting to study the structure of these
algebraic equations.  Even more generally, the pseudo-potential
(\ref{eq:9.3}) at $x=0$ may be replaced by a linear superposition of a
finite number of such pseudo-potentials at $x=x_1$, $x_2$, \dots.  The
number of coefficients in the solution increases but remains finite,
leading to more simultaneous algebraic equations that are still linear. 
The Green's functions can be treated in a very similar manner. 

Needless to say, the range of integration in Eq.\ (\ref{eq:9.1}) for $x'$
can be replaced by a semi-infinite or finite interval, and the $V(x,x')$
may contain additional terms such as those from step potentials.  A more
interesting problem is to apply the Fermi pseudo-potentials to
first-order differential equations.

In summary, the theory of the Fermi pseudo-potential in one dimension has
been worked out here together with the simplest non-trivial application
to a problem of current interest.

\begin{acknowledgments}
For helpful discussions, we are indebted to Maurice Jacob, Harold
Levine, Andr\'e Martin, John Myers, and Raymond Stora.  One of us (TTW)
thanks the Theoretical Physics Division of CERN for their kind
hospitality.  Some of the thinking in this paper has been discussed in
an unpublished Harvard preprint by John Myers and TTW. 
\end{acknowledgments}       

\appendix 

\section{}\label{app:1} 

A possible way to solve Eq.\ (\ref{eq:3.13}) for $F(\kappa)$ is as
follows.  Because of the exponential function in the last term, let 
\begin{equation}
x=e^{F(\kappa)}\label{eq:A1}
\end{equation}
and
\begin{equation}
y=\ln \kappa.\label{eq:A2}
\end{equation}
[This $x$ of Eq.\ (\ref{eq:A1}) of course has nothing to do with the
space variable in the Schr\"odinger equation (\ref{eq:1.1}), for
example.]  In view of Eq.\ (\ref{eq:3.3}), Eq.\ (\ref{eq:3.13}) is
translationally invariant in $y$.  It is therefore desirable to use $x$
as the independent variable and $y$ as the dependent variable, leading to
a first-order ordinary differential equation for
\begin{equation}
z=\frac{dy}{dx}.\label{eq:A3}
\end{equation}
This first-order equation is
\begin{equation}
-\frac{2}{x^2z^3}\left[x\,\frac{dz}{dx}+z\right] = \frac{1}{x^2z^2}-1
+(c_3^2 +c_2c_4)x^2.\label{eq:A4}
\end{equation}
Let
\begin{equation}
u=\frac{1}{x^3z^2}.\label{eq:A5}
\end{equation}
Then, after some algebra,
\begin{equation}
\frac{du}{dx}=-\frac{1}{x^2}+(c_3^2+c_2c_4).\label{eq:A6}
\end{equation}
Integration yields
\begin{equation}
u=\frac{1}{x}+2c_1+(c_3^2+c_2c_4)x,\label{eq:A7}
\end{equation}
where $c_1$ is the fourth arbitrary constant of integration.  The
expression for $z$ follows from Eqs.\ (\ref{eq:A5}) and (\ref{eq:A7}):
\begin{equation}
z=\pm x^{-1}[1+2c_1x+(c_3^2+c_2c_4)x^2]^{-1/2},\label{eq:A8}
\end{equation}
where the $\pm$ sign comes from taking the square root of $z^2$.  It is,
of course, related to the fact that every term in Eq.\ (\ref{eq:A4}) is
even in $z$.  In the following, there are many $\pm$ and $\mp$ signs; it
is to be understood that these signs are used as a shorthand for two
equations, one with the upper sign everywhere and a second one with the
lower sign everywhere.

From Eqs.\ (\ref{eq:A3}) and (\ref{eq:A8}), $\,y$ is given by
\begin{equation}
y=\mp \int \,\frac{dx'}{\sqrt{(c_3^2+c_2c_4)+2c_1x'+x'{}^2}},\label{eq:A9}
\end{equation}
where $x'=1/x$.  It is fortunate that this integral is elementary.  For
definiteness, consider the case
\begin{equation}
c_3^2+c_2c_4-c_1^2>0.\label{eq:A10}
\end{equation}
In this case, the explicit integration of the right-hand side of Eq.\
(\ref{eq:A9}) gives
\begin{equation}
y=\mp\left[\sinh^{-1}\frac{1+c_1x}{\sqrt{c_3^2+c_2c_4-c_1^2x}}
+\mbox{const.}\right].\label{eq:A11}
\end{equation}
The rest of the calculation is straightforward although lengthy, and the
results are
\begin{eqnarray}
f_1(\kappa)&=&\frac{-\sqrt{c_3^2+c_2c_4-c_1^2}\,[c_0\kappa
+(c_0\kappa)^{-1}]\mp 2c_3}{\sqrt{c_3^2+c_2c_4-c_1^2}\,[c_0\kappa
-(c_0\kappa)^{-1}]\pm 2c_1},\label{eq:A12}\\[4pt]
f_2(\kappa)&=&1\mp \frac{2c_2}{\sqrt{c_3^2+c_2c_4-c_1^2}\,[c_0\kappa
-(c_0\kappa)^{-1}]\pm 2c_1},\label{eq"A13}\\[4pt]
f_3(\kappa)&=&\frac{-\sqrt{c_3^2+c_2c_4-c_1^2}\,[c_0\kappa
+(c_0\kappa)^{-1}]\pm 2c_3}{\sqrt{c_3^2+c_2c_4-c_1^2}\,[c_0\kappa
-(c_0\kappa)^{-1}]\pm 2c_1},\label{eq:A14}\\[4pt]
f_4(\kappa)&=&1\mp \frac{2c_4}{\sqrt{c_3^2+c_2c_4-c_1^2}\,[c_0\kappa
-(c_0\kappa)^{-1}]\pm 2c_1}.\label{eq:A15}
\end{eqnarray}
Since $c_1$, $c_2$, $c_3$ and $c_4$ are arbitrary constants of
integration, their signs can be changed simultaneously.  This change then
removes all the $\pm$ and $\mp$ signs, and Eqs.\
(\ref{eq:A12})--(\ref{eq:A15}) reduce to Eqs.\ (\ref{eq:3.14}).

The other case where $c_3^2+c_2c_4-c_1^2<0$ can be treated in an entirely
similar manner, leading to Eq.\ (\ref{eq:3.16}). 

\section{}\label{app:B}

In this Appendix B, the reasoning is given that leads to Eq.\
(\ref{eq:7.2}). 

Just as the $\Psi^{\text{in}}$ of Eq.\ (\ref{eq:7.1}) contains information
about the initial state of the quantum memory and the behavior of the
incoming wave $\psi^{\text{in}}$ before scattering, the $\Psi^{\text{out}}$
contains information about the final state of the quantum memory and the
behavior of the outgoing wave after scattering.  Later the outgoing
wave moves away from the memory and the information about this outgoing
wave is no longer available.  This means that the final state of the
memory is given by $\Psi^{\text{out}}$ \textit{averaged} over this outgoing
wave.  This average can be written schematically as
\begin{equation}
M=\int \Psi^{\text{out}}\Psi^{\text{out}\dag}.\label{eq:B1}  
\end{equation}
This $M$ is the density matrix for the quantum memory.  Here the $\int$
indicates integration and summation over all degrees of freedom
associated with the outgoing wave or particle, but not those of the
quantum memory.  The corresponding differential symbol is omitted:  It
is $d^3\textbf{r}$ if the wave is described by the three-dimensional
Schr\"odinger equation; it is $d^3\textbf{r}$ together with a summation
over the spin in the case of the Dirac equation; and it is a functional
differential such as $\mathcal{D}A_{\mu}$ in the context of quantum
field theory.

In order for the quantum memory to function, the final state must be a
pure state $\sum_j a_j^{\text{out}}|j\rangle$, just like the initial
state.  Therefore, the above $M$ must also be given by
\begin{equation}
M=\sum_j a_j^{\text{out}}|j\rangle\,\sum_i a_i^{\text{out}*}\langle
i|.\label{eq:B2}
\end{equation}
Equating the two formulas for $M$ gives
\begin{equation}
\sum_j a_j^{\text{out}}|j\rangle\,\sum_i a_i^{\text{out}*}\langle i| = \int
\Psi^{\text{out}}\Psi^{\text{out}\dag}.\label{eq:B3}
\end{equation}
It remains to derive Eq.\ (\ref{eq:7.2}) from Eq.\ (\ref{eq:B3}).

Since the $|j\rangle$ form a complete set, $\,\Psi^{\text{out}}$ can always
be written in the form
\begin{equation}
\Psi^{\text{out}}=\sum_j |j\rangle\,\psi_j^{\text{out}}.\label{eq:B4}
\end{equation}
The substitution of Eq.\ (\ref{eq:B4}) into Eq.\ (\ref{eq:B3}) gives
\begin{equation}
a_j^{\text{out}}a_i^{\text{out}*}=\int \psi_j^{\text{out}}
\psi_i^{\text{out}\dag}\label{eq:B5}
\end{equation}
for all $i$ and $j$.  For $i\ne j$, define the integral
\begin{equation}
I_{ji}=\int (a_i^{\text{out}}\psi_j^{\text{out}} -a_j^{\text{out}}
\psi_i^{\text{out}})(a_i^{\text{out}*}\psi_j^{\text{out}\dag}
-a_j^{\text{out}*}
\psi_i^{\text{out}\dag}).\label{eq:B6}
\end{equation}
This $I_{ji}$ is non-negative, and is zero only if
\begin{equation}
a_i^{\text{out}}\psi_j^{\text{out}}-a_j^{\text{out}}\psi_i^{\text{out}}
=0.\label{eq:B7}
\end{equation} 
But the substitution of Eq.\ (\ref{eq:B5}) into Eq.\ (\ref{eq:B6}) gives
immediately that
\begin{equation}
I_{ji}=0.\label{eq:B8}
\end{equation}
Thus, Eq.\ (\ref{eq:B7}) holds for all $i$ and $j$.

The desired Eq.\ (\ref{eq:7.2}) is just Eq.\ (\ref{eq:B4}) with Eq.\
(\ref{eq:B7}).

\end{document}